\documentclass[prd,twocolumn,nofootinbib,showpacs,amsmath,amssymb,floatfix,eqsecnum]{revtex4-1}

\usepackage{graphicx,color,hyperref}
\usepackage{times}

\begin{document}

\def \beq {\begin{equation}}
\def \eeq {\end{equation}}
\def \beqa {\begin{eqnarray}}
\def \eeqa {\end{eqnarray}}
\def \bseq {\begin{subequations}}
\def \eseq {\end{subequations}}
\newcommand \dg {\dagger}
\newcommand \up {\uparrow}
\newcommand \down {\downarrow}
\newcommand \al {\alpha}
\newcommand \be {\beta}
\newcommand \sg {\sigma}
\newcommand \ran {\rangle}
\newcommand \lan {\langle}
\newcommand \un {\underline}
\newcommand \ep {\epsilon}
\newcommand \lam {\lambda}
\newcommand \pd {\partial}
\newcommand \pdb {\bar{\partial}}
\newcommand \mb {\mathbf}
\newcommand \ms {\boldsymbol}
\newcommand \yhat {\mathbf{\hat{y}}}
\newcommand \tb {\bar{t}}
\newcommand \nnb {\nonumber}
\newcommand \D {\delta}
\newcommand \T {\hat{T}}
\newcommand \K {\hat{K}}
\newcommand \II {\mathbb{I}}
\newcommand \gamt {\tilde{\gamma}}
\newcommand \gam {\gamma}
\newcommand \dt {\tilde{d}}
\newcommand \Om {\Omega}
\newcommand \J {\mathcal{J}}
\newcommand \vphi {\varphi}
\newcommand \XX {\mathbb{X}}
\newcommand \A {\mathcal{A}}
\newcommand \vth {\vartheta}

\title{Canonical quantization of nonlinear sigma models with theta term, with applications to symmetry-protected topological phases}

\author{Matthew F. Lapa}
\email[email address: ]{lapa2@illinois.edu}
\author{Taylor L. Hughes}
\affiliation{Department of Physics and Institute for Condensed Matter Theory, University of Illinois at Urbana-Champaign, 61801-3080}

\date{\today}

\begin{abstract}

We canonically quantize $O(D+2)$ nonlinear sigma models (NLSMs) with theta term on arbitrary smooth, closed, connected, oriented 
$D$-dimensional spatial manifolds $\mathcal{M}$, with the goal of proving the suitability of these models for describing 
symmetry-protected topological (SPT) phases of bosons in $D$ spatial dimensions. We show that in the disordered phase of
the NLSM, and when the coefficient $\theta$ of the theta term is an integer multiple of $2\pi$, the theory on $\mathcal{M}$ has a 
unique ground state and a finite energy gap to all excitations. We also construct the ground state wave functional of the 
NLSM in this parameter regime, 
and we show that it is independent of the metric on $\mathcal{M}$ and given by the exponential
of a \emph{Wess-Zumino} term for the NLSM field, in agreement with previous results on flat space. Our results
show that the NLSM in the disordered phase and at $\theta=2\pi k$, $k\in\mathbb{Z}$, has a symmetry-preserving ground
state but no topological order (i.e., no topology-dependent ground state degeneracy), making it an ideal model for describing
SPT phases of bosons. Thus, our work places previous results on SPT phases derived using NLSMs on solid theoretical ground.
To canonically quantize the NLSM on $\mathcal{M}$ we use Dirac's method for the quantization of systems with second class 
constraints, suitably modified to account for the curvature of space.  In a series of four appendices we provide the  
technical background needed to follow the discussion in the main sections of the paper. 

\end{abstract}


\maketitle

\section{Introduction}
\label{sec:intro}

Nonlinear sigma models (NLSMs), quantum field theories in which the field is a map from spacetime to a \emph{target manifold}
$\mathcal{T}$, have a long history of study in both high-energy and condensed matter 
physics~\cite{gell1960axial,BP1975,polyakov1975interaction,brezin1976spontaneous,brezin1976renormalization,hamer1979strong,
feynman1981qualitative,Haldane1983,friedan1985nonlinear,ShankarRead,AbanovWiegmann,SenthilFisher,
xu2013nonperturbative}. The earliest example of such a model was introduced in Ref.~\cite{gell1960axial} by Gell-Mann and 
L\'{e}vy and applied to the theory of $\beta$-decay. 
Some time later these models were brought to the attention of condensed matter physicists when, for example, Haldane showed that the 
$O(3)$ NLSM with \emph{theta term} and coefficient (``theta angle")  $\theta=2\pi S$ described the continuum 
limit of a spin-$S$ Heisenberg chain in one spatial dimension~\cite{Haldane1983}. 
The theta term is a particular topological term that can be added to the NLSM action when the dimension of the target
manifold $\mathcal{T}$ is the same as the dimension of spacetime. 
Very recently, interest in NLSMs with theta term has experienced a 
revival due to the proposal, formalized in Ref.~\cite{CenkeClass1}, that the disordered phase of an $O(D+2)$ 
NLSM with theta term and $\theta=2\pi k$, $k\in\mathbb{Z}$, can describe (a subset of) symmetry-protected topological 
(SPT) phases of bosons in $D$ spatial dimensions (for general references on SPT phases we refer the reader to 
Refs.~\cite{gu2009tensor,pollmann2010entanglement,chen2012symmetry,levin2012braiding,chen2013symmetry,senthil2015review}).
The work of many authors has provided a trove of evidence supporting this 
proposal~\cite{CenkeClass2,CenkeBraiding,CenkeLine,CenkeSU2N,CenkeDual,xu2013wave,ElseNayak,liu2013symmetry,
you2016stripe,you2016decorated,lapa2016topological}. However, despite the many successes of the NLSM description, several 
outstanding issues still require clarification. In particular, the ground state of the $O(D+2)$ NLSM, in the 
parameter regime relevant for the description of SPT phases, has only been studied on flat space~\cite{xu2013wave}. A thorough 
study of the ground state (or states) of this theory, as well as the energy gap to the first excited states, should be carried out on 
\emph{arbitrary} curved spatial manifolds (with various topologies) in order to establish the suitability of the $O(D+2)$ NLSM with theta term 
(in the disordered phase and at $\theta=2\pi k$, $k\in\mathbb{Z}$) as a model of bosonic SPT 
phases. For example, it is important to check that the model has a unique ground state no matter the underlying spatial manifold, as befits a 
system without topological order. It is the purpose of this paper to provide such a study. 

%

Let us begin by providing the setup for the description of bosonic SPT phases in $D$ spatial dimensions using $O(D+2)$ 
NLSMs~\cite{CenkeClass1}.
The target manifold of the $O(D+2)$ NLSM is $\mathcal{T}=S^{D+1}$, the unit $(D+1)$-sphere, and the NLSM field is a 
$(D+2)$-component unit vector field $\mb{n}$ with components $n^a$, $a=1,\dots,D+2$. If the SPT phase is protected by a symmetry 
group $G$, then the symmetry transformation information is naturally encoded in the NLSM description of that phase 
via a homomorphism $\sigma: G \to SO(D+2)$, which assigns to each group element $g\in G$ an $SO(D+2)$ matrix $\sigma(g)$ 
which rotates the NLSM field $\mb{n}$\footnote{For symmetries which also have a 
nontrivial action on spacetime, for example time-reversal symmetry, the homomorphism instead takes the form 
$\sigma: G \to O(D+2)$. In other words, for symmetries with a nontrivial action on spacetime one must allow for the possibility of 
orientation-reversing transformations on the target space of the NLSM.}. 
The reason for mapping $G$ into $SO(D+2)$ is that the $O(D+2)$ NLSM \emph{with} 
theta term has an $SO(D+2)$ global symmetry, and so embedding $G$ inside $SO(D+2)$ immediately guarantees the invariance of the NLSM 
description of the SPT phase under the action of $G$, at least at the classical level. At the quantum level the symmetry is 
expected to be unbroken only in the disordered phase of the NLSM, which is the phase of interest for the description of SPT phases.
 For a given group $G,$ many distinct homomorphisms $\sigma$ are possible, 
and the different possibilities (modulo a notion of ``independent NLSMs" explained in Ref.~\cite{CenkeClass1}) correspond to 
different SPT phases with the symmetry of the group $G$.

The NLSM description has been shown to capture many of the physical properties of bosonic SPT phases. For example, it can predict the
structure of the ground state wave functional~\cite{xu2013wave,Tanaka2016} 
and the braiding statistics of point and loop-like excitations in gauged SPT 
phases~\cite{CenkeBraiding}. In addition, in Ref.~\cite{ElseNayak} an explicit connection was made between the NLSM 
classification of SPT phases of Refs.~\cite{CenkeClass1,CenkeClass2}, and the group cohomology classification of bosonic SPT 
phases of Ref.~\cite{chen2013symmetry}. Very recently, the present authors demonstrated how the NLSM description of 
bosonic SPT phases can be combined with the theory of gauged Wess-Zumino actions to compute the topological electromagnetic response of 
some bosonic SPT phases in all dimensions~\cite{lapa2016topological}. 

All of these works strongly support the idea that the $O(D+2)$
NLSM in its disordered phase, with theta term and $\theta=2\pi k$, $k\in\mathbb{Z}$, and a suitable symmetry 
assignment $\sigma: G\to SO(D+2)$, can describe nontrivial SPT phases with symmetry group $G$. However, 
several of the defining properties of an SPT phase are based on the behavior of the SPT phase when it is placed on (closed) spatial 
manifolds $\mathcal{M}$ of arbitrary topology, and the NLSM description of SPT phases has not yet been tested in this setting. 
To be precise, let $\mathcal{M}$ be an arbitrary smooth, 
closed\footnote{A \emph{closed} manifold is a compact manifold without boundary.}, connected, oriented, 
$D$-dimensional manifold. We also equip $\mathcal{M}$ with a Riemannian metric.
In this paper we prove the following three properties of the $O(D+2)$ NLSM when formulated on spatial manifolds $\mathcal{M}$ of
this kind.
\begin{enumerate}
\item The ground state of the $O(D+2)$ NLSM in the disordered phase and at $\theta=2\pi k$, $k\in\mathbb{Z}$, on
$\mathcal{M}$ is \emph{unique}.
\item The ground state wave functional of the NLSM on $\mathcal{M}$ is \emph{independent} of the metric on $\mathcal{M}$, and is 
proportional to the exponential of a suitably-defined \emph{Wess-Zumino} term for the NLSM field $\mb{n}$, just as in the case on flat
space~\cite{xu2013wave}.
\item There is a finite energy gap between the ground state and the first excited state of the $O(D+2)$ NLSM (in the disordered 
phase and at $\theta=2\pi k$, $k\in\mathbb{Z}$) on $\mathcal{M}$. 
\end{enumerate}
These three properties together imply that the $O(D+2)$ NLSM in the disordered phase and at $\theta=2\pi k$, 
$k\in\mathbb{Z}$, represents a system with SPT order, but \emph{not} topological order (no topology-dependent ground
state degeneracy). In particular, the fact that the ground state wave functional involves a Wess-Zumino term for $\mb{n}$
implies that the ground state is invariant under the action of the group $G$ which protects the SPT phase. This is equivalent
to the statement that the ground state does not spontaneously break the symmetry of the group $G$, which is a crucial property
of an SPT phase (see, for example, the discussion in the introduction of Ref.~\cite{levin2012braiding}).

In order to prove these statements we canonically quantize the $O(D+2)$ NLSM with theta term on $(D+1)$-dimensional spacetimes of the 
form $\mathcal{M}\times \mathbb{R}$, where $\mathcal{M}$ is a $D$-dimensional spatial manifold and $\mathbb{R}$ represents time. As 
stated above, we assume that the spatial manifold $\mathcal{M}$ is smooth, closed, connected, and oriented, and we equip
$\mathcal{M}$ with a Riemannian metric. The canonical quantization of the $O(N)$ NLSM on flat space and for various $N$, in 
various dimensions, and with various topological terms has been considered previously in
Refs.~\cite{corrigan1979non,feynman1981qualitative,maharana1983quantization,bowick1986fractional,
olczyk1990canonical,karabali1991soliton}. In particular, Ref.~\cite{olczyk1990canonical} considered
the $O(3)$ NLSM with theta term in one spatial dimension, and Refs.~\cite{bowick1986fractional,karabali1991soliton} 
considered the $O(3)$ NLSM with \emph{Hopf term} in two spatial dimensions. To carry out the quantization these references used 
Dirac's method~\cite{Dirac,HRT} for the quantization of systems with second class constraints, 
and we follow the same route in this paper (with suitable modifications to account for the curvature of the space $\mathcal{M}$). 
This formalism is necessary to handle the constraint that the NLSM field $\mb{n}$ be a unit vector field, which is equivalent to the 
statement that the target space $\mathcal{T}$ of the $O(D+2)$ NLSM is the unit sphere $S^{D+1}$. After the quantization
we study this theory in its disordered limit, and in that limit we prove the three properties of this model which are stated above. 


This paper is organized as follows. In Sec.~\ref{sec:NLSM-quantization} we introduce the NLSM with theta term and discuss the 
canonical quantization of this model on flat space. In Sec.~\ref{sec:wavefcn-gap} we compute the ground state wave functional and 
the energy gap of the NLSM in the disordered phase at $\theta=2\pi k$, $k\in\mathbb{Z}$ on flat space. 
In Sec.~\ref{sec:NLSM-quantization-curved} we quantize the NLSM on curved spaces $\mathcal{M}$, and then compute the ground 
state wave functional and the energy gap, as well as prove the uniqueness of the ground state of the theory on $\mathcal{M}$. 
Sec.~\ref{sec:conclusion} presents our conclusions. The paper also includes several appendices containing additional background 
material necessary for the discussion in the main sections of the paper. In Appendix~\ref{app:NLSM-one-dim} we 
review the solution of the quantum mechanics problem of a particle constrained to the surface of a sphere $S^{N-1}$, equivalent to 
the $O(N)$ NLSM in one spacetime dimension. In Appendix~\ref{app:reg} we explain the need for regularization of the NLSM 
Hamiltonian, and we also discuss an alternative regularization scheme from the one used in Sec.~\ref{sec:wavefcn-gap}. In 
Appendix~\ref{app:symplectic} we review the symplectic geometry approach to the Hamiltonian mechanics of a continuum field 
theory. Finally, in Appendix~\ref{app:WZ-intrinsic} we explain a simple example of an intrinsic construction of a Wess-Zumino term 
for the NLSM field  on $\mathcal{M}$ which does not require the use of a higher-dimensional manifold $\mathcal{B}$ which has 
$\mathcal{M}$ as its boundary.

\section{Nonlinear sigma models, Hamiltonian formalism, and canonical quantization}
\label{sec:NLSM-quantization}

In this section we introduce the NLSM with theta term and discuss its canonical quantization on flat
space $\mathbb{R}^D$ using Dirac's method for quantization in the presence of constraints. We show in some detail that this system 
possesses only two second class constraints, regardless of the value of the coefficient $\theta$ of the theta term. 
We then compute the classical \emph{Dirac brackets} for the
NLSM using the two second class constraints. Finally, following Dirac's prescription, we obtain the commutation relations for the quantum 
theory from the Dirac brackets in the same way that one obtains the commutation relations from the Poisson brackets of an unconstrained 
classical theory. We also discuss a functional Schroedinger representation of these 
commutation relations, previously used in a field theory context in Ref.~\cite{karabali1991soliton}, which we use throughout the paper 
for concrete calculations.

\subsection{NLSM and theta term}

The $O(N)$ NLSM in $D+1$ spacetime dimensions is a theory of an $N$-component vector field 
$\mb{n}$ with components $n^a$, $a= 1,\dots, N$, subject to the constraint $\mb{n}\cdot\mb{n}= n_a n^a= 1$ 
(so $\mb{n}$ is a \emph{unit} vector field). The action for the NLSM takes the form
\beq
	S[\mb{n}]= \int d^{D+1} x\ \frac{1}{2f}(\pd^{\mu}n_a)(\pd_{\mu} n^a)\ ,  
\eeq
where $x^{\mu}$, $\mu= 0,\dots,D$, ($x^0 = t$) are the spacetime coordinates, $d^{D+1}x= dx^0\cdots dx^D$, and 
we sum over all indices (Latin or Greek) which appear once as a superscript and once as a subscript. 
Also, we use the notation $\pd_{\mu}\equiv\frac{\pd}{\pd x^{\mu}}$. 
Latin indices are raised and lowered using the metric $\delta_{ab}$ and its inverse $\delta^{ab}$, 
while Greek indices are raised and lowered using
the ``mostly minus" Minkowski metric $\eta_{\mu\nu}$ (i.e., as a matrix $\eta = \text{diag}(1,-1,\dots,-1)$) and its inverse
$\eta^{\mu\nu}$. We use units in which $c=\hbar=1$. 
In this section of the paper we work on flat, $(D+1)$-dimensional Minkowksi spacetime, denoted by
$\mathbb{R}^{D,1}$. Finally, the quantity $f$ is a positive coupling constant with units of (length)$^{D-1}$ (so that the bare 
NLSM field is dimensionless). For $D>1$ the model is in an ordered phase for small $f$ and a disordered phase for 
large $f$\footnote{For $D>1$ the coupling constant $f$ has units so the magnitude of the coupling constant can only be 
established with respect to a reference scale $f^*$. Such an $f^*$ is naturally provided by the location of the zero of the beta 
function $\beta(f)$ of the coupling constant $f$ for $D>1$ (Ref.~\cite{hamer1979strong} showed that there is no 
zero in $D=1$). The value $f=f^*$ is the point at which the renormalization group flow of $f$ 
crosses over from a flow towards the ordered phase to a flow towards the disordered phase at low energies.}. 
For $D=1$ there is no ordered phase, and the renormalization group flow at any scale is always 
towards the disordered phase~\cite{hamer1979strong}.

The target manifold $\mathcal{T}$ of the $O(N)$ NLSM is the unit sphere $S^{N-1}$ and,
in the special case that $N= D+2$, the dimension of the 
target manifold is the same as the dimension of spacetime. For this particular value of $N$ there is an interesting topological term, 
called the theta term, which can be added to the action for the NLSM. This topological term is simply the pullback to spacetime of the 
volume form on $S^{D+1}$ via the map $\mb{n}: \mathbb{R}^{D,1} \to S^{D+1}$. We now explain this in more detail.

Let $\omega_{D+1}$ be the volume form on $S^{D+1}$ (with the radius of the sphere set to one). If the sphere is 
parametrized by the coordinates $n^a$, $a=1,\dots,D+2$, subject to the constraint $n^a n_a=1$, then the volume
form in these coordinates is
\beq
	\omega_{D+1}= \sum_{a=1}^{D+2} (-1)^{a-1} n^a dn^1 \wedge \cdots \wedge \overline{dn^a}\wedge \cdots\wedge dn^{D+2}\ , \label{eq:volume-form}
\eeq
where the overline means to omit that term from the wedge product. 
In what follows we also use the notation 
$\mathcal{A}_{D+1} \equiv \text{Area}[S^{D+1}]=\frac{2\pi^{\frac{D+2}{2}}}{\Gamma(\frac{D+2}{2})}$ 
for the area of $S^{D+1}$ (so $\mathcal{A}_1=2\pi$, $\mathcal{A}_2= 4\pi$, etc.).
With this notation the theta term can be written compactly as
\beq
	 S_{\theta}[\mb{n}]= \frac{1}{\mathcal{A}_{D+1}}\int_{\mathbb{R}^{D,1}} \mb{n}^*\omega_{D+1}\ ,
\eeq
where $\mb{n}^*\omega_{D+1}$ denotes the pullback to spacetime of the form $\omega_{D+1}$ via the map 
$\mb{n}: \mathbb{R}^{D,1}\to S^{D+1}$. In coordinates the theta term can be written as
\beq
	S_{\theta}[\mb{n}]= \frac{1}{\mathcal{A}_{D+1}}\int d^{D+1}x\ \ep_{a_1\cdots a_{D+2}} n^{a_1}\pd_{t}n^{a_2}\pd_{1}n^{a_3}\cdots\pd_{D}n^{a_{D+2}}\ . \label{eq:theta-term-coords}
\eeq

The full action for the NLSM with theta term takes the form
\beq
	S[\mb{n}]= \int d^{D+1} x\ \frac{1}{2f}(\pd^{\mu}n_a)(\pd_{\mu} n^a) - \theta S_{\theta}[\mb{n}]\ ,
\eeq
where the dimensionless parameter $\theta$ is the theta angle discussed in Sec.~\ref{sec:intro}. The Lagrangian which
follows from this action is
\beq
	\mathcal{L}= \frac{1}{2f}(\pd^{\mu}n_a)(\pd_{\mu} n^a) + \frac{\theta}{\A_{D+1}}B_a(\pd_t n^a)\ ,
\eeq
where we introduced
\beq
	B_a = \ep_{a a_1\cdots a_{D+1}}n^{a_1}\pd_{1}n^{a_2}\cdots\pd_{D}n^{a_{D+1}}\ . \label{eq:Ba}
\eeq
At this point we are now in a position to proceed with the canonical quantization of this system.

\subsection{Quantization of constrained systems}

Due to the constraint $n^a n_a=1$ on the the NLSM field, the canonical quantization of the NLSM requires Dirac's theory of 
constrained Hamiltonian systems, and the use of Dirac brackets in 
particular~\cite{Dirac}. Let us briefly sketch, following Ref.~\cite{HRT}, the steps involved in the quantization of a 
constrained system. We first recall some basic definitions. A constraint $\phi$ is a function on phase space which is to be set equal 
to zero. Two functions $f$ and 
$g$ on phase space are \emph{strongly equivalent} if they are equal throughout phase space. This is just ordinary equality of
functions, $f=g$. Two functions $f$ and $g$ are called \emph{weakly equivalent} if they become equal when
all constraints $\phi$ are set to zero. Weak equivalence of two functions $f$ and $g$ is denoted by $f \approx g$\footnote{In this paper
the symbol ``$\approx$''  is \emph{only} used to denote weak equivalence.}.

The first step in the quantization of a constrained system is to find all of the constraints. 
This step involves the construction of a modified Hamiltonian $\tilde{H}$ 
such that the time derivative of any constraint $\phi$, as
given by the modified Hamiltonian, weakly vanishes. In other words, we have $\frac{d}{dt}\phi = \{\phi,\tilde{H}\} \approx 0$,
where $\{\cdot,\cdot\}$ denotes the ordinary Poisson bracket. This is a consistency condition on the time evolution of the 
dynamical system, as constraints should not change with time. In general the modified Hamiltonian $\tilde{H}$ is distinct from the 
original Hamiltonian $H$ obtained from the Lagrangian via a Legendre transformation.

The next step in the quantization is to isolate the \emph{second class constraints}, and then use these
constraints to construct the \emph{Dirac bracket}. Recall that the second class constraints are those constraints which have 
non-vanishing Poisson brackets with each other.  Let $\psi_I(\mb{x})$, $I= 1,\dots,N_c$, denote the second class constraints in 
our system\footnote{In a field theory the constraints are usually functions of the position $\mb{x}$ in space.}, and define the 
functions $M_{IJ}(\mb{x},\mb{y})$ by
\beq
	M_{IJ}(\mb{x},\mb{y})= \{\psi_I(\mb{x}),\psi_J(\mb{y})\}\ . \label{eq:def-M-IJ}
\eeq
The functions $M_{IJ}(\mb{x},\mb{y})$ should be viewed as the components of
a matrix $\mb{M}$ with discrete indices $I,J$ and continuous spatial indices $(\mb{x},\mb{y})$. 
In terms of this matrix the Dirac bracket for two functions
$f(\mb{x})$ and $g(\mb{y})$ on phase space is given by
\begin{widetext}
\beq
	 \{f(\mb{x}),g(\mb{y})\}_{D} = \{f(\mb{x}),g(\mb{y})\} - \sum_{I,J=1}^{N_c} \int d^D\mb{z}\ d^D\mb{z}'\ \{f(\mb{x}),\psi_I(\mb{z})   \}M^{-1}_{IJ}(\mb{z},\mb{z}')\{\psi_J(\mb{z}'), g(\mb{y})\}\ , \label{eq:Dirac-bracket}
\eeq
\end{widetext}
where the functions $M^{-1}_{IJ}(\mb{x},\mb{y})$ are the components of the inverse matrix to $\mb{M}$ in the
sense that
\beq
	\sum_{J=1}^{N_c}\int d^D\mb{y}\ M_{IJ}(\mb{x},\mb{y}) M^{-1}_{JK}(\mb{y},\mb{z})= \delta_{IK}\delta^{(D)}(\mb{x}-\mb{z})\ .
\eeq

Finally, the equal-time commutators in the quantized theory are obtained from the classical Dirac brackets according to the
rule
\beq
	\{f(\mb{x}),g(\mb{y})\}_D \to  -i[ \hat{f}(\mb{x}),\hat{g}(\mb{y})]\ , \label{eq:bracket-to-commutator}
\eeq
where $\hat{f}(\mb{x})$ is the operator in the quantum theory corresponding to the classical function $f(\mb{x})$. The
quantum theory is then defined by the operator $\hat{H}$ corresponding to the \emph{original} Hamiltonian $H$ obtained from the 
Legendre transformation of the Lagrangian (not the modified Hamiltonian $\tilde{H}$ discussed above), 
combined with the equal-time commutators obtained from the classical Dirac brackets\footnote{By construction, in the
quantum theory defined in this way one has $[\hat{\tilde{H}},\hat{\mathcal{O}}] \approx [\hat{H},\hat{\mathcal{O}}]$ for any 
operator $\hat{\mathcal{O}}$. Thus, the operator $\hat{H}$ corresponding to the original Hamiltonian may be used in the
quantum theory instead of the operator $\hat{\tilde{H}}$ corresponding to the modified Hamiltonian.}. 

\subsection{Canonical quantization of the $O(D+2)$ NLSM with theta term}

We now carry out the program outlined in the last section for the NLSM with theta term. To start, the momentum conjugate to 
$n^a$ is
\beq
	\pi_a\equiv \frac{\pd \mathcal{L}}{\pd (\pd_t n^a)}= \frac{1}{f}(\pd_t n_a) + \frac{\theta}{\A_{D+1}}B_a \ , \label{eq:momentum}
\eeq
and the Hamiltonian obtained from the Lagrangian via the Legendre transformation 
$H= \int d^D\mb{x}\left( \pi_a \pd_t n^a - \mathcal{L}\right)$ is
\beq
	H= \int d^D \mb{x}\ \left\{ \frac{f}{2}\left(\pi_a - \frac{\theta}{\A_{D+1}}B_a\right)^2 + \frac{1}{2f}(\nabla n^a)^2 \right\}\ , \label{eq:NLSM-ham}
\eeq
where $\mb{x}= (x^1,\dots,x^D)$ is the vector of spatial coordinates, $d^D \mb{x}= \prod_{j=1}^D dx^{j}$, $\nabla$ is the 
spatial gradient, and $(\nabla n^a)^2 \equiv (\nabla n^a)\cdot(\nabla n_a)$, etc. For this system the Poisson bracket of any
two functionals $F_1$ and $F_2$ of the fields $n^a(\mb{x})$ and their conjugate momenta $\pi_a(\mb{x})$ is given by
\beq
	\{F_1,F_2\}= \int d^D\mb{x} \left( \frac{\delta F_1}{\delta n^a(\mb{x})}\frac{\delta F_2}{\delta \pi_a(\mb{x})}- \frac{\delta F_1}{\delta \pi_a(\mb{x})}\frac{\delta F_2}{\delta n^a(\mb{x})}  \right)\ ,
\eeq
where $\frac{\delta}{\delta n^a(\mb{x})}$ is a functional derivative. 

We now move on to the problem of finding all of the constraints for this system. To begin with we have only the single constraint
\beq
	\psi_1(\mb{x})= n_a(\mb{x}) n^a(\mb{x}) - 1\ . \label{eq:constraint1}
\eeq
Setting $\psi_1(\mb{x})= 0$ enforces the condition that $\mb{n}$ is a unit vector field. Following Dirac's procedure we now 
use this constraint to construct a modified Hamiltonian $\tilde{H}$ such that $\{\psi_1(\mb{x}),\tilde{H}\}\approx 0$. 
As a first attempt towards the construction of $\tilde{H}$, we define the modified Hamiltonian $H'$ by
\beq
	H'= H + \int d^D\mb{y}\ u_1(\mb{y})\psi_1(\mb{y})\ ,
\eeq
where $u_1(\mb{x})$ is an as yet undetermined function. Note that $H'\approx H$ since the constraint weakly vanishes,
$\psi_1(\mb{x})\approx 0$. Using the product rule for the Poisson bracket, we find that
\begin{align}
	\{\psi_1(\mb{x}),H'\}&= \{\psi_1(\mb{x}),H\} \\
 + \int d^D\mb{y}&\ \big( \{\psi_1(\mb{x}),u_1(\mb{y})\}\psi_1(\mb{y}) + u_1(\mb{y})\{\psi_1(\mb{x}),\psi_1(\mb{y})\} \big)\ . \nnb
\end{align}
A short computation shows that $\{\psi_1(\mb{x}),\psi_1(\mb{y})\}=0$. Then, since the constraint $\psi_1(\mb{y})$ is weakly equivalent
to zero, we find that
\beq
	\{\psi_1(\mb{x}),H'\}\approx \{\psi_1(\mb{x}),H\}\ .
\eeq
Finally, due to the identity $n^a(\mb{x})B_a(\mb{x})= 0$, we have $\{\psi_1(\mb{x}),H\}= 2 f\ n^a(\mb{x})\pi_a(\mb{x})$
for any value of $\theta$. This means that 
\beq
	\{\psi_1(\mb{x}),H'\}\approx 2 f\ n^a(\mb{x})\pi_a(\mb{x})\ ,
\eeq
and so we find a second constraint
\beq
	\psi_2(\mb{x})= n^a(\mb{x})\pi_a(\mb{x})\ ,
\eeq
which must also be set to zero for consistent time evolution of this system.

We now make a further modification to the Hamiltonian and define
\beq
	H''= H + \sum_{I=1}^2\int d^D\mb{y}\ u_I(\mb{y})\psi_I(\mb{y})\ ,
\eeq
where we introduced a second undetermined function $u_2(\mb{y})$, 
and investigate the conditions under which $\{\psi_2(\mb{x}),H''\}\approx 0$. After some algebra we find that
\beq
	\{\psi_2(\mb{x}),H''\}\approx \{\psi_2(\mb{x}),H\} + \int d^D\mb{y}\ u_1(\mb{y})\{\psi_2(\mb{x}),\psi_1(\mb{y})\}\ .
\eeq
At this point it is possible that, depending on the form of $\{\psi_2(\mb{x}),\psi_1(\mb{y})\}$, the equation
\beq
	\{\psi_2(\mb{x}),H\} + \int d^D\mb{y}\ u_1(\mb{y})\{\psi_2(\mb{x}),\psi_1(\mb{y})\} = 0 \label{eq:solve-u1}
\eeq
can be solved to yield a function $u_1(\mb{x})$ such that $\{\psi_2(\mb{x}),H''\}\approx 0$. Below we show that this is
indeed the case. This means that the constraints $\psi_1$ and $\psi_2$ account for all of the constraints in this problem, and
it also means that the additional function $u_2(\mb{x})$ is not needed for the construction of the modified Hamiltonian. 
Therefore we can set $u_2(\mb{x})= 0$ at this point. However, we note here that the fact that $u_2(\mb{x})$ can be 
set to zero is specific to this particular problem. It is easy to imagine a scenario in which the function $u_2(\mb{x})$ would not be zero, for
example if the Poisson bracket of $\psi_1$ and $\psi_2$ were to vanish, then requiring $\{\psi_2(\mb{x}),H''\}\approx 0$ would yield
a new constraint $\psi_3(\mb{x})= \{\psi_2(\mb{x}),H\}$. If that were the case, then it is very likely that we would need a non-zero 
$u_2(\mb{x})$ to construct a final modified Hamiltonian $\tilde{H}$ such that $\{\psi_3(\mb{x}),\tilde{H}\} \approx 0$.
However, for the problem considered here we can
safely set $u_2(\mb{x})=0$, and so we find that the final modified Hamiltonian is given by
\beq
	\tilde{H}=  H + \int d^D\mb{y}\ u_1(\mb{y})\psi_1(\mb{y})\ ,
\eeq
where $u_1(\mb{x})$ solves Eq.~\eqref{eq:solve-u1}. Finally, we note that for the purposes of the 
quantization we do not need to know the exact form of $\tilde{H}$. This is fortunate because the Poisson bracket 
$\{\psi_2(\mb{x}),H\}$ is fairly complicated in the case that the parameter $\theta$ is non-zero.

Now that we know all of the constraints in the problem, we can look at their Poisson brackets with each other. For this
system we find that the function $M_{IJ}(\mb{x},\mb{y})$ defined in Eq.~\eqref{eq:def-M-IJ} has the explicit form
\beq
	M_{IJ}(\mb{x},\mb{y})= 2i (\sigma^y)_{IJ}\ r^2(\mb{x})\delta^{(D)}(\mb{x}-\mb{y})\ ,
\eeq
where $(\sigma^y)_{IJ}$ is the $(I,J)$ element of the second Pauli matrix $\sigma^y$, and where we defined the 
radial coordinate 
\beq
	r^2(\mb{x}) =  n_a(\mb{x})n^a(\mb{x})\ .
\eeq
In terms of $r^2(\mb{x})$ the first second class constraint in the NLSM problem reads as $\psi_1(\mb{x})= r^2(\mb{x})-1$. 
The inverse of $M_{IJ}(\mb{x},\mb{y})$ is
\beq
	M^{-1}_{IJ}(\mb{x},\mb{y})= -\frac{i}{2} (\sigma^y)_{IJ}\frac{1}{r^2(\mb{x})}\delta^{(D)}(\mb{x}-\mb{y})\ . \label{eq:M-inv}
\eeq
The fact that the inverse exists means that Eq.~\eqref{eq:solve-u1} can indeed be solved for the function $u_1(\mb{x})$,
as we claimed above.

We can now use the components $M^{-1}_{IJ}(\mb{x},\mb{y})$ from Eq.~\eqref{eq:M-inv} to construct the classical Dirac brackets for this 
theory (see Eq.~\eqref{eq:Dirac-bracket} for the definition of the Dirac bracket). To quantize the theory we then replace all functions 
$f(\mb{x})$ with operators $\hat{f}(\mb{x})$ and replace the Dirac brackets with commutators as shown in 
Eq.~\eqref{eq:bracket-to-commutator}. After following these steps, we find that the equal-time commutation relations for the $O(D+2)$ 
NLSM with theta term, \emph{for any value} of $\theta$, are given by
\begin{widetext}
\begin{subequations}
\label{eq:comm-rels-NLSM-1st-form}
\beqa
	\left[\hat{n}^a(\mb{x}), \hat{n}^b(\mb{y})\right] &=& 0  \\
	\left[\hat{n}^a(\mb{x}), \hat{\pi}_b(\mb{y})\right] &=& i\left({\delta^a}_b - \frac{\hat{n}^a(\mb{x}) \hat{n}_b(\mb{y})}{\hat{r}^2(\mb{x})}\right)\delta^{(D)}(\mb{x}-\mb{y}) \\
	\left[\hat{\pi}_a(\mb{x}), \hat{\pi}_b(\mb{y})\right] &=& \frac{i}{\hat{r}^2(\mb{x})}(\hat{\pi}_a(\mb{x}) \hat{n}_b(\mb{y}) - \hat{\pi}_b(\mb{y}) \hat{n}_a(\mb{x}))\delta^{(D)}(\mb{x}-\mb{y})\ . 
\eeqa
\end{subequations}
\end{widetext}
At this point we note that by construction $\hat{r}^2(\mb{x})$ commutes with $\hat{\pi}_a(\mb{y})$ (in fact it commutes
with any operator on the Hilbert space). Therefore $\hat{r}^2(\mb{x})$ is in the center of the algebra defined by the
commutation relations in Eqs.~\eqref{eq:comm-rels-NLSM-1st-form} and it is consistent to plug in the constraint 
$\hat{r}^2(\mb{x})=1$. An explicit discussion about this point can be found in Ref.~\cite{kleinert1997proper} in 
the context of the quantum mechanical problem of a free particle constrained to move on the surface of a sphere. The fact
that $\hat{r}^2(\mb{x})$ commutes with any operator on the Hilbert space simply follows from the fact that the Dirac bracket
of any functional $F$ on phase space with a constraint $\psi_I(\mb{x})$ is \emph{strongly equal} to zero, $\{F,\psi_I(\mb{x})\}_D=0$.
From now on we work with the commutation relations obtained after this substitution. These have the form
\begin{widetext}
\begin{subequations}
\label{eq:comm-rels-NLSM}
\beqa
	\left[\hat{n}^a(\mb{x}), \hat{n}^b(\mb{y})\right] &=& 0  \\
	\left[\hat{n}^a(\mb{x}), \hat{\pi}_b(\mb{y})\right] &=& i({\delta^a}_b - \hat{n}^a(\mb{x}) \hat{n}_b(\mb{y}))\delta^{(D)}(\mb{x}-\mb{y}) \\
	\left[\hat{\pi}_a(\mb{x}), \hat{\pi}_b(\mb{y})\right] &=& i(\hat{\pi}_a(\mb{x}) \hat{n}_b(\mb{y}) - \hat{\pi}_b(\mb{y}) \hat{n}_a(\mb{x}))\delta^{(D)}(\mb{x}-\mb{y})\ , \label{eq:momentum-commutator}
\eeqa
\end{subequations}
\end{widetext}
and they have appeared in several papers on the canonical quantization of the $O(N)$
NLSM~\cite{corrigan1979non,maharana1983quantization,bowick1986fractional,olczyk1990canonical,karabali1991soliton}.
However, we emphasize that we have explicitly shown here
that these commutators are valid for the NLSM with theta term \emph{for any value} of the parameter $\theta$. 
We also note that there is an operator ordering ambiguity in the commutation relation for two momenta, however, we can say 
that the two terms on the right-hand side of Eq.~\eqref{eq:momentum-commutator} should have the \emph{same} ordering,
so that the commutator has the important property that
$\left[\pi_a(\mb{x}), \pi_b(\mb{y})\right] = -\left[\pi_b(\mb{y}), \pi_a(\mb{x})\right]$.

To make progress in analyzing the NLSM we employ a functional Schroedinger representation of the commutation relations
of Eqs.~\eqref{eq:comm-rels-NLSM} in which $\hat{n}^a(\mb{x})$ acts as multiplication by the function $n^a(\mb{x})$, and
$\hat{\pi}_a(\mb{x})$ is given in terms of a functional derivative with respect to $n^a(\mb{x})$ as
\beq
	\hat{\pi}_a(\mb{x})= -i \left({\delta_a}^b - n_a(\mb{x})n^b(\mb{x})\right)\frac{\delta}{\delta n^b(\mb{x})}\ . \label{eq:schro-rep}
\eeq
This choice reproduces \emph{all} of the commutators shown in Eqs.~\eqref{eq:comm-rels-NLSM} 
(with the operator ordering indicated there). This Schroedinger representation was used previously in
Ref.~\cite{karabali1991soliton} to construct soliton operators in the $O(3)$ NLSM with Hopf term in three spacetime 
dimensions. It has also been used in the study of the $O(N)$ NLSM in one spacetime 
dimension~\cite{podolsky1928quantum,dewitt1957dynamical,kleinert1997proper,neto1999does,neves2000stuckelberg,
hong2000improved,abdalla2001quantisation,scardicchio2002classical,hong2004gauged}, which is equivalent
to the quantum mechanics problem of a free particle in $\mathbb{R}^N$ confined to the surface of the sphere $S^{N-1}$.
In Appendix~\ref{app:NLSM-one-dim} we review the solution of this quantum mechanical model using this Schroedinger
representation. We use the results of Appendix~\ref{app:NLSM-one-dim} 
in Sec.~\ref{sec:wavefcn-gap} and Sec.~\ref{sec:NLSM-quantization-curved} 
to study the energy gap in the $O(D+2)$ NLSM in the limit of infinitely large coupling $f$ on flat and 
curved space, respectively. 

\section{Ground state wave functional and the energy gap on flat space}
\label{sec:wavefcn-gap}

In this section we study the $O(D+2)$ NLSM with theta term in the disordered ($f\to\infty$) phase with $\theta= 2\pi k$, 
$k\in\mathbb{Z}$, on \emph{flat} space $\mathbb{R}^D$.
We give an alternative derivation, within the canonical formalism, of the result of Ref.~\cite{xu2013wave}
for the ground state wave functional of the NLSM at these parameter values. Finally,
we use a lattice regularization of the NLSM to prove the uniqueness of the ground state and the existence of an energy gap
in the disordered phase of the model at $\theta=2\pi k$. This section should be viewed as a warm up for 
Sec.~\ref{sec:NLSM-quantization-curved} in which we discuss the ground state wave functional and the energy gap of the NLSM on 
an arbitrary spatial manifold $\mathcal{M}$.

\subsection{Ground state wave functional at large $f$ and $\theta= 2\pi k$, $k\in\mathbb{Z}$}

We first discuss the construction of the ground state wave functional. As discussed above, we consider the
disordered phase of the model in which $f\to\infty$. In this limit the Hamiltonian operator is approximately given by
\beq
	\hat{H} = \frac{f}{2}\int d^D \mb{x}\ \left(\hat{\pi}_a - \frac{\theta}{\A_{D+1}}\hat{B}_a\right)^2\ ,
\eeq
where we ignore terms proportional to $\frac{1}{f}$.
Since the Hamiltonian in this limit is expressed as an integral over space of the square of a local operator, the lowest possible
energy of any eigenstate is zero. This means that the ground state wave functional of the NLSM in this limit is
determined only by the property that it is annihilated by the operators 
\beq
	\hat{\mathcal{D}}^{(\theta)}_a= \hat{\pi}_a - \frac{\theta}{\A_{D+1}}\hat{B}_a\ ,\ a= 1,\dots,D+2\ .
\eeq
On the other hand, because of the specific form of the operator $\hat{\pi}^a(\mb{x})$ in the Schroedinger representation 
Eq.~\eqref{eq:schro-rep} used in this paper, the Hamiltonian needs to be regularized in some way before any excited states can be 
constructed. We discuss the need for regularization of the Hamiltonian in more detail in Appendix~\ref{app:reg}. 
For now, however, we are only interested in the construction of the ground state wave functional, and so we can delay the issue of 
regularization of the Hamiltonian until the next subsection.

To start, consider the case where $\theta=0$. Since $\hat{\pi}_a$ is proportional to a functional derivative with respect to 
$n^a$, and since the energy is bounded below by zero, we can see that the ground state wave functional is just 
a constant, $\Psi_{\theta=0}[\mb{n}]= 1$. 
A general state $|\Psi\ran$ in the Hilbert space of the NLSM can be expanded in the ``position basis"
$\{|\mb{n}\ran\}$, which contains a state $|\mb{n}\ran$ for every possible configuration of the NLSM field on the space $\mathbb{R}^D$.
The field operator $\hat{n}^a(\mb{x})$ is diagonal in this basis, $\hat{n}^a(\mb{x})| \mb{n}\ran = n^a(\mb{x})| \mb{n}\ran$, where
$n^a(\mb{x})$ is the \emph{function} corresponding to the particular state $|\mb{n}\ran$. 
For a more precise formulation we should restrict the set $\{|\mb{n}\ran\}$ to include only those field configurations on $\mathbb{R}^D$
with finite (classical) potential energy. This restriction implies a choice of boundary condition on the field configurations at spatial infinity,
for example we could choose $\mb{n}(\mb{x}) \to \mb{n}_0$, a particular constant field configuration, as $|\mb{x}|\to\infty$. 

A general state in this basis takes the form
\beq
	|\Psi\ran \propto \int [D\mb{n}] \Psi[\mb{n}]| \mb{n}\ran\ ,
\eeq
where $\Psi[\mb{n}]$ is the wave functional (i.e., the amplitude of the basis state $|\mb{n}\ran$ in the full state $|\Psi\ran$), 
and the integration is over all possible configurations of the field $\mb{n}$ at every point in space (possibly subject to 
a boundary condition at spatial infinity ensuring finite energy). We define the 
measure $ [D\mb{n}]$ to be the product over all points $\mb{x}$ in space of the volume form $\omega_{D+1}$ on the sphere $S^{D+1}$.
Since the ground state wave functional of the NLSM at $\theta=0$ is just $\Psi_{\theta=0}[\mb{n}]= 1$, it follows that the state vector for 
the ground state is just an equal weight superposition of all basis states,
\beq
	|\Psi_{\theta=0}\ran \propto \int [D\mb{n}] | \mb{n}\ran\ .
\eeq
This state can be thought of as a continuum analogue of a trivial paramagnetic state.

Next, we look at the ground state for non-zero $\theta$ in the particular case that $\theta$ is an integer multiple of
$2\pi$. In this case it is possible to remove the term $\frac{\theta}{\A_{D+1}}\hat{B}_a$
from the operator $\hat{\mathcal{D}}^{(\theta)}_a$ 
via a well-defined unitary transformation, which means that the ground state 
in this case can be obtained by multiplying the ground state at $\theta=0$ by a unitary operator. As we discussed
in Sec.~\ref{sec:intro}, the case $\theta=2\pi k$, $k\in\mathbb{Z}$, is also interesting from a physical point of view because
for these values of $\theta$ the $O(D+2)$ NLSM has been shown to capture many of the physical properties of SPT
phases of bosons in $D$ spatial 
dimensions~\cite{CenkeClass1,CenkeClass2,CenkeBraiding,CenkeLine,CenkeSU2N,CenkeDual,xu2013wave,ElseNayak,liu2013symmetry,
you2016stripe,you2016decorated,lapa2016topological}.

The ground state wave functional at $\theta=2\pi k$ can be constructed using a Wess-Zumino (WZ) term for the NLSM field
$\mb{n}$. Recall now that we are working in $(D+1)$-dimensional Minkowski spacetime $\mathbb{R}^{D,1}$, so that the
physical space is just $\mathbb{R}^D$. The WZ term is written as an integral over the extended space 
$\mathcal{B}= [0,1]\times \mathbb{R}^D$, where $\mathbb{R}^D$
represents the original $D$-dimensional space, and $[0,1]$ is an auxiliary direction of space
used in the construction of the WZ term. We use the notation $\tilde{n}^a(\mb{x},s)$ to denote the
extension of the NLSM field $n^a(\mb{x})$ into the extra direction, where $\mb{x}\in\mathbb{R}^D$ and $s\in[0,1]$.
Typically one chooses boundary conditions in the
auxiliary direction so that $\tilde{n}^a(\mb{x},0)= {\delta^a}_{D+2}$ (i.e., a trivial configuration) and 
$\tilde{n}^a(\mb{x},1)= n^a(\mb{x})$, so that the physical space sits at $s=1$.

We now show that for $\theta=2\pi k$, $k\in\mathbb{Z}$, the ground state wave functional is
\beq
	\Psi_{\theta=2\pi k}[\mb{n}]= e^{-ikS_{WZ}[\mb{n}]}\ , \label{eq:ground-state-k}
\eeq
where $S_{WZ}[\mb{n}]$ is the WZ term,
\begin{widetext}
\beqa
	S_{WZ}[\mb{n}] &=& \frac{2\pi}{\A_{D+1}}\int_{\mathcal{B}}\tilde{\mb{n}}^*\omega_{D+1} \nnb \\
	&=& \frac{2\pi}{\A_{D+1}}\int_0^1 ds \int d^D\mb{x}\ \ep_{a_1\cdots a_{D+2}} \tilde{n}^{a_1}\pd_{s}\tilde{n}^{a_2}\pd_{1}\tilde{n}^{a_3}\cdots\pd_{D}\tilde{n}^{a_{D+2}}\ .
\eeqa
The WZ term involves the pullback of the volume form $\omega_{D+1}$ on $S^{D+1}$ to the extended
space $\mathcal{B}$ via the map $\tilde{\mb{n}}: \mathcal{B}\to S^{D+1}$.
To prove Eq.~\eqref{eq:ground-state-k} we first recall the formula for the variation of the WZ term,
\beq
	\delta S_{WZ}[\mb{n}] = -\frac{2\pi}{\A_{D+1}}\int d^D\mb{x}\ \ep_{a_1\cdots a_{D+2}} \delta n^{a_1} n^{a_2}\pd_{1} n^{a_3}\cdots\pd_{D}n^{a_{D+2}}\ , \label{eq:WZ-variation}
\eeq
which is an integral only over the physical space $\mathbb{R}^D$ (in the case that we can neglect terms coming from
the boundary of physical space). Then we have
\beqa
	\frac{\delta}{\delta n^a(\mb{x})}\Psi_{\theta=2\pi k}[\mb{n}] &=& i \frac{2\pi k}{\A_{D+1}} \left( \ep_{a a_2\cdots a_{D+2}}  n^{a_2} \pd_{1} n^{a_3}\cdots\pd_{D} n^{a_{D+2}}       \right)  \Psi_{\theta=2\pi k}[\mb{n}] \nnb \\
	&=&  i \frac{\theta}{\A_{D+1}} B_a(\mb{x}) \Psi_{\theta=2\pi k}[\mb{n}] \ .
\eeqa
\end{widetext}
Then, using $\hat{\pi}_a(\mb{x})= -i ({\delta_a}^b - n_a(\mb{x})n^b(\mb{x}))\frac{\delta}{\delta n^b(\mb{x})}$ and the 
fact that $n^b B_b = 0$, we find that
\beq
	\hat{\mathcal{D}}^{(\theta=2\pi k)}_a \Psi_{\theta=2\pi k}[\mb{n}]= 0\ ,
\eeq
which completes the proof. The state vector for the ground state at $\theta=2\pi k$ then takes the form
\beq
	|\Psi_{\theta=2\pi k}\ran \propto \int [D\mb{n}] e^{-ikS_{WZ}[\mb{n}]} |\mb{n}\ran\ .
\eeq
Thus, we have succeeding in re-deriving the result of Ref.~\cite{xu2013wave} for the ground state wave functional of
this system within the canonical formalism.

The relationship between the ground state wave functionals at $\theta=0$ and $\theta=2\pi k$ can be understood
in terms of a unitary transformation of the Hamiltonian by the \emph{operator}
\beq
	\hat{\mathcal{U}}^{(k)}=  e^{-ikS_{WZ}[\hat{\mb{n}}]}\ . \label{eq:unitary-operator}
\eeq
In the Schroedinger representation, and using a suitable test functional, one can show that
\beqa
	\hat{\mathcal{U}}^{(k),\dg} \hat{\mathcal{D}}^{(\theta=2\pi k)}_a \hat{\mathcal{U}}^{(k)} &=& \hat{\mathcal{D}}^{(\theta=0)}_a \nnb \\
	&=& \hat{\pi}_a\ ,
\eeqa
which means that
\beq
	\hat{\mathcal{U}}^{(k),\dg} \hat{H}_{(\theta=2\pi k)} \hat{\mathcal{U}}^{(k)} = \hat{H}_{(\theta=0)}\ , \label{eq:unitary-trans}
\eeq
and that
\beq
	|\Psi_{\theta=2\pi k}\ran= \hat{\mathcal{U}}^{(k)} |\Psi_{\theta=0}\ran\ . \label{eq:states-0-2pik}
\eeq
Note also that since $\hat{\mathcal{U}}^{(k)}$ commutes with the potential energy term $\frac{1}{2f}(\nabla n^a)^2$,  
Eq.~\eqref{eq:unitary-trans} holds for the full Hamiltonian of Eq.~\eqref{eq:NLSM-ham} (i.e., not just in the large $f$ limit). In
fact, \emph{for any values} of $f$ and $\theta$ the full Hamiltonian obeys the relation
\beq
	\hat{\mathcal{U}}^{(1),\dg} \hat{H}_{(\theta)} \hat{\mathcal{U}}^{(1)} = \hat{H}_{(\theta-2\pi)}\ ,
\eeq
which shows that the spectrum of the $O(D+2)$ NLSM with theta term is $2\pi$-periodic in the value of the parameter
$\theta$. This is a crucial result since it will let us simultaneously study the energy spectra for any values of $\theta$ related by a $2\pi$ shift.

We see that the theta angle of the NLSM enters into the Hamiltonian of Eq.~\eqref{eq:NLSM-ham} 
as something like a gauge field, and the derivative 
operator $\hat{\mathcal{D}}^{(\theta)}_a$ looks like a covariant derivative. In the case that $\theta=2\pi k$, $k\in\mathbb{Z}$,
we can interpret the phase of the ground state wave functional as being obtained from a gauge transformation which removes the 
``gauge field" term $\frac{\theta}{\A_{D+1}}\hat{B}_a$ from $\hat{\mathcal{D}}^{(\theta)}_a$ at the expense of 
an additional phase in the wave functional. This gauge transformation, however, can only be performed when 
$\theta$ is an integer multiple of $2\pi$. This is because the exponential $e^{-ikS_{WZ}[\mb{n}]}$ of the WZ term, which involves 
an extension of the field $n^a$ into an auxiliary direction, is only well-defined (i.e., independent of the extension) when 
$k$ is an integer~\cite{witten1983global}. To be precise, we note here that to apply the argument of 
Ref.~\cite{witten1983global} on the quantization of $k$ we must replace flat space $\mathbb{R}^D$ with a 
$D$-dimensional sphere so that space is a compact manifold (the radius of the sphere can be taken to be very large so that the
curvature is nearly zero). The original infinite space $\mathbb{R}^D$ is then obtained in the limit that the radius of the $D$-sphere 
goes to infinity. We now move on to a discussion of the uniqueness of the ground state and the calculation of the energy gap
in the NLSM at $\theta=2\pi k$ and $f\to \infty$. 

\subsection{Uniqueness of the ground state and the energy gap at large $f$ and $\theta=2\pi k$, $k\in\mathbb{Z}$}

In the previous subsection we showed that the NLSM Hamiltonians at $\theta=2\pi k$ and $\theta=0$ are related by a unitary 
transformation, which means that the energy spectrum in this model at $\theta=2\pi k$ is identical to the spectrum at $\theta=0$. 
In the context of applications to SPT phases, one of the most important properties of the NLSM at large $f$ that we would like to
verify is the uniqueness of the ground state and the existence of an energy gap between the 
ground state and all of the excited states. In this subsection we use a lattice regularization of the NLSM at large $f$
to prove the uniqueness of the ground state and the existence of an energy gap at $\theta=0$. 
Since the NLSM Hamiltonian at $\theta=2\pi k$ is related to the Hamiltonian at $\theta=0$ by a unitary transformation, 
the uniqueness of the ground state and the existence of an energy gap at $\theta=2\pi k$ follow immediately from this
result at $\theta=0$. In Appendix~\ref{app:reg}
we also present an alternative regularization procedure for the NLSM Hamiltonian in the disordered limit, and we show that this
alternative procedure gives a result for the energy gap which is consistent with the result derived in this section using a
lattice regularization. Therefore we expect that our result for the energy gap of the $O(D+2)$ NLSM in its disordered limit
is independent of the specific details of the regularization scheme used in the calculation.


To start we consider a hypercubic lattice with spacing $a$ and coordinates which are vectors with integer entries and
denoted by boldface Latin letters $\mb{j},\mb{k}$, etc. 
The continuum coordinate $\mb{x}$ is given in terms of the lattice coordinate $\mb{j}$ by $\mb{x}= a\mb{j}$. In
the lattice regularization the Dirac delta function is represented by 
$\delta^{(D)}(\mb{x}-\mb{y}) \to \frac{1}{a^D}\delta_{\mb{j}\mb{k}}$ if $\mb{x}= a\mb{j}$ and $\mb{y}= a\mb{k}$. 
If we define lattice operators $\hat{n}^a_{\mb{j}}$ and $\hat{\pi}_{a,\mb{j}}$ by
\bseq
\beqa
	\hat{n}^a(a\mb{j}) &=& \hat{n}^a_{\mb{j}} \\
	\hat{\pi}_a(a\mb{j}) &=& \frac{1}{a^D}\hat{\pi}_{a,\mb{j}}\ ,
\eeqa
\eseq
where $\hat{n}^a(a\mb{j})$ and $\hat{\pi}_a(a\mb{j})$ are the continuum field operators at $\mb{x}=a\mb{j}$, 
then the NLSM commutation relations of Eqs.~\eqref{eq:comm-rels-NLSM} become
\bseq
\beqa
	\left[\hat{n}^a_{\mb{j}}, \hat{n}^b_{\mb{k}}\right] &=& 0  \\
	\left[\hat{n}^a_{\mb{j}}, \hat{\pi}_{b,\mb{k}}\right] &=& i({\delta^a}_b - \hat{n}^a_{\mb{j}} \hat{n}_{b,\mb{k}})\delta_{\mb{j}\mb{k}} \\
	\left[\hat{\pi}_{a,\mb{j}}, \hat{\pi}_{b,\mb{k}}\right] &=& i(\hat{\pi}_{a,\mb{j}} \hat{n}_{b,\mb{k}} - \hat{\pi}_{b,\mb{k}} \hat{n}_{a,\mb{j}})\delta_{\mb{j}\mb{k}}\  .
\eeqa
\eseq
The integration over space becomes $\int d^D\mb{x} \to a^D \sum_{\mb{j}}$, and so the regularized Hamiltonian at large
$f$ and $\theta=0$ takes the form
\beq
	\hat{H}(a) = \frac{f}{2 a^D}\sum_{\mb{j}}\hat{\pi}_{a,\mb{j}}{\hat{\pi}^a}_{\mb{j}}\ . \label{eq:reg-Ham}
\eeq
Here we have written $\hat{H}(a)$ to indicate the explicit dependence of the Hamiltonian on the cutoff $a$.

The regularized Hamiltonian Eq.~\eqref{eq:reg-Ham} is a sum of many identical Hamiltonians for an $O(N)$ NLSM in
one spacetime dimension, with $N=D+2$. In Appendix~\ref{app:NLSM-one-dim} we review the solution of this quantum 
mechanics problem using Dirac's formalism for quantizing constrained systems. Using the results from
Appendix~\ref{app:NLSM-one-dim} we can rewrite the Hamiltonian as 
\beq
	\hat{H}(a) = \frac{f}{2 a^D}\sum_{\mb{j}}\hat{\mathcal{C}}_{\mb{j}}\ ,
\eeq
where $\hat{\mathcal{C}}_{\mb{j}}$ is the quadratic Casimir of $so(D+2)$ formed from the conserved charge operators
in the $O(D+2)$ NLSM on site $\mb{j}$. We immediately deduce that the \emph{unique} ground state of this system is the state
with
\beq
	\hat{\mathcal{C}}_{\mb{j}} = 0\ ,\ \forall\ \mb{j}\ ,
\eeq
i.e., the state which is the tensor product of the trivial representation of $SO(D+2)$ on all sites $\mb{j}$. The energy gap, which
is equal to the energy of the first excited state, is (``m" stands for mass)
\beq
	m(a)= \frac{f}{2 a^D}(D+1)\ .\label{eq:mass-gap-flat-space}
\eeq
This energy corresponds to the case that one site in the lattice is excited to a state in the fundamental representation of 
$SO(D+2)$.  In the theory at large $f$ the first excited state is highly degenerate, but this degeneracy will be broken by the 
inclusion of a small kinetic energy term (with coefficient $\frac{1}{f}$), which will cause the energies of all degenerate states in the 
first excited state manifold to disperse. 

We see that for a fixed bare coupling constant $f$, the energy gap $m(a)$ goes to infinity as we take the continuum limit 
$a\to 0$. On the other hand, it is more physical to make the coupling constant cutoff-dependent, $f\to f(a)$, and demand
that $f(a)$ depend on the cutoff $a$ in such a way as to make the mass gap $m(a)$ independent of the cutoff $a$ used
to define the theory. Following the procedure of Ref.~\cite{hamer1979strong}, we demand that $\frac{d m(a)}{da}=0$, 
which yields the renormalization group equation for $f(a)$ in the regime of large $f$,
\beq
	a \frac{d f(a)}{da}= D f(a)\ . \label{eq:RG-eqn-f}
\eeq
We find that $f(a) \to \infty$ in the infra-red (i.e., low energy) limit $a\to \infty$, which confirms the validity of our expansion of the 
Hamiltonian in powers of $\frac{1}{f}$. Integrating Eq.~\eqref{eq:RG-eqn-f} from some reference scale $a_0$ in the ultra-violet, at 
which $f=f_0$, to the scale $a$, we find that $f(a)$ is given in terms of $f_0$ as $f(a)= \left(\frac{a}{a_0}\right)^D f_0$, so that 
the mass gap $m$ (which is now independent of $a$)  is given in terms of $f_0$ and the reference scale $a_0$ by
\beq
	m= \frac{f_0}{2 a_0^D}(D+1)\ .
\eeq

\section{Quantization, ground state wave functional, and energy gap on curved space}
\label{sec:NLSM-quantization-curved}

In this section we repeat the analysis of Secs.~\ref{sec:NLSM-quantization} and \ref{sec:wavefcn-gap} in the case that the 
spacetime takes the form $\mathcal{M}\times\mathbb{R}$, where $\mathbb{R}$ represents the time direction 
and $\mathcal{M}$ is a curved, $D$-dimensional manifold representing space (the precise assumptions on the properties
of $\mathcal{M}$ were stated in Sec.~\ref{sec:intro} and are repeated below). In particular, we will accomplish
the goal of the paper, which is to prove the three properties of the NLSM on curved space which are stated in Sec.~\ref{sec:intro}. 
That is, we prove the uniqueness of the ground state and the existence of an energy gap in the 
$O(D+2)$ NLSM in the disordered ($f\to\infty$) phase at $\theta=2\pi k$, $k\in\mathbb{Z}$, on arbitrary spatial manifolds 
$\mathcal{M}$, and we also explicitly construct the ground state wave functional on $\mathcal{M}$. We find that the wave 
functional takes the form of an exponential of a WZ term for $n^a$, just as in the case on flat space~\cite{xu2013wave}. 
To prove the uniqueness of the ground state and the existence of an energy gap in the NLSM on $\mathcal{M}$, 
we use a \emph{triangulation} of the manifold to set up a lattice-like regularization of the NLSM Hamiltonian at large $f$. Within
this regularization scheme, the demonstration of the uniqueness of the ground state and the computation of the energy
gap can be done in a way which is very similar to the calculation on flat space from Sec.~\ref{sec:wavefcn-gap}. 

The results of this section prove that the $O(D+2)$ NLSM, in the parameter regime studied in this paper, possesses SPT order,
but not topological order, and is therefore a suitable model for SPT phases. 
One interesting aspect of the theory on a curved space $\mathcal{M}$ is that for certain choices of manifold $\mathcal{M}$
the standard construction of the WZ term fails, and so alternative constructions are needed. We discuss the standard
construction of the WZ term  and one type of alternative construction in some detail in 
this section. Then, in Appendix~\ref{app:WZ-intrinsic} we give an explicit example of a third construction which can be
used when the other two constructions fail.  Before we discuss these details, however, we need to first explain the
modifications to the canonical quantization procedure of Sec.~\ref{sec:NLSM-quantization} which are needed to study the
NLSM in the canonical formalism on the curved space $\mathcal{M}$.

\subsection{Canonical quantization of the NLSM on a curved space}

In this subsection we discuss the canonical quantization of the $O(D+2)$ NLSM with theta term on a spacetime of the form
$\mathcal{M}\times\mathbb{R}$, where $\mathbb{R}$ represents the time direction 
and $\mathcal{M}$ is a smooth, closed, connected, oriented, $D$-dimensional manifold. 
We take the metric on spacetime to have the form (we use a ``mostly minus" signature for the metric)
\beq
	g= dt\otimes dt - G_{ij}(\mb{x})dx^i\otimes dx^j\ , \label{eq:metric}
\eeq
where $i,j=1,\dots,D$ (and a sum over repeated indices is implied). On flat Minkowski space we have $G_{ij}(\mb{x})= \delta_{ij}$, but in 
the general case $G_{ij}(\mb{x})$ are the components of a Riemannian metric on $\mathcal{M}$. In addition, we have
\beq
	\text{det}[g]= (-1)^D\text{det}[G]\ .
\eeq
By a common abuse of notation we will also use the letters $g$ and $G$ to denote $\text{det}[g]$ and 
$\text{det}[G]$, respectively, for the remainder of the paper.

To start, we use the formalism of Appendix~\ref{app:symplectic} to understand how to quantize a free scalar field on a curved 
space. The key piece of information we need is the appropriate form of the Poisson bracket for a free scalar field on a curved space. With this 
information in hand we can then use Dirac's procedure to quantize the NLSM on a curved space, since the $O(D+2)$ NLSM consists 
of $D+2$ scalar fields, but subject to the additional constraint $n^a(\mb{x})n_a(\mb{x})=1$. At this point we suggest
that the reader skim through Appendix~\ref{app:symplectic} to understand our notation for the symplectic geometry approach to
studying field theories in the Hamiltonian formalism. 

First, we outline our general strategy for determining the correct symplectic form $\Omega$ to use to describe a field theory
on a curved space. Suppose that the system we would like to study on a curved space has a definition in terms of an action 
\beqa
	S &=& \int d^{D+1}x\ \sqrt{(-1)^Dg}\ \mathcal{L} \nnb \\
	&=&\int d^{D+1}x\ \sqrt{G}\mathcal{L}\ ,
\eeqa
where $\mathcal{L}$ is the Lagrangian and we assumed a metric on spacetime of the form of Eq.~\eqref{eq:metric}. 
In this case our strategy for determining the appropriate symplectic form is to choose $\Omega$ such that the Hamilton equations 
of motion obtained from $\Omega$ via Eq.~\eqref{eq:Hamilton-eqns}
coincide with the Euler-Lagrange equations of motion obtained from the action for our system on a curved space. Once
we know the correct $\Omega$, we can use it to find the correct Poisson brackets from Eq.~\eqref{eq:PB}. These
Poisson brackets will then give us the information we need to find the commutation relations for the fields in the
quantum field theory on the curved space $\mathcal{M}$.

Let us see how this all plays out in the case of a free scalar field $\phi$. In this case the Lagrangian is
\beqa
	\mathcal{L} &=& \frac{1}{2}(\pd^{\mu}\phi)(\pd_{\mu}\phi) \nnb \\
	&=&  \frac{1}{2}\left[(\pd_t\phi)^2 - G^{ij}\pd_i\phi\pd_j\phi\right]\ ,
\eeqa
where in the first line $\mu= 0,1,\dots,D$ (and $x^0= t$). In the second line we specialized to the case of curved
space only (i.e., a metric of the form shown in Eq.~\eqref{eq:metric}), and we used the tensor $G^{ij}$ which satisfies the relation
$G^{ij}G_{jk}= {\delta^i}_k$. The momentum conjugate to $\phi$ is $\pi = \frac{\pd \mathcal{L}}{\pd (\pd_t\phi)}= \pd_t\phi$, 
and the Hamiltonian is
\beqa
	H &=& \int d^D\mb{x}\ \sqrt{G}\left( \pi \pd_t\phi - \mathcal{L} \right) \nnb \\
	&=& \frac{1}{2}\int d^D\mb{x}\ \sqrt{G} \left( \pi^2 + G^{ij}\pd_i\phi\pd_j\phi\right)\ .
\eeqa
Starting from the action $S= \int d^{D+1}\ \sqrt{G}\mathcal{L}$, we can derive the Euler-Lagrange equation of motion for $\phi$,
\beq
	\pd_t^2\phi - \frac{1}{\sqrt{G}}\pd_i\left(\sqrt{G} G^{ij}\pd_j\phi\right) = 0\ . \label{eq:Euler-Lagrange-EOM}
\eeq

Now that we know the Euler-Lagrange equation of motion for $\phi$, we can look for a choice of $\Omega$ so that the Hamilton 
equations obtained from it are equivalent to this Euler-Lagrange equation. We find that the choice of $\Omega$ which yields the
correct equations of motion is
\beq
	\Omega= \int d^D\mb{x}\ \sqrt{G(\mb{x})}\ \delta\pi(\mb{x})\wedge\delta\phi(\mb{x})\ .
\eeq
Indeed, using this form of $\Omega$ with Eq.~\eqref{eq:Hamilton-eqns} we find that
\begin{subequations}
\beqa
	\pd_t\phi &=& \pi \\
	\pd_t\pi &=& \frac{1}{\sqrt{G}}\pd_i\left(\sqrt{G} G^{ij}\pd_j\phi\right) \ ,
\eeqa 
\end{subequations}
which is clearly equivalent to the equation of motion Eq.~\eqref{eq:Euler-Lagrange-EOM} derived from the action. 
Using the correct form of $\Omega$ we can
now derive the form of the Poisson bracket for $\phi$ and $\pi$ on curved space. First, using Eq.~\eqref{eq:vector-field}
we find that the vector fields on the phase space corresponding to the functionals $\phi(\mb{x})$ and $\pi(\mb{x})$ are
\begin{subequations}
\beqa
	\un{V}_{\phi(\mb{x})} &=& -\frac{1}{\sqrt{G(\mb{x})}}\frac{\delta}{\delta\pi(\mb{x})} \\
	\un{V}_{\pi(\mb{x})} &=& \frac{1}{\sqrt{G(\mb{x})}}\frac{\delta}{\delta\phi(\mb{x})}\ .
\eeqa	
\end{subequations}
From these we find that
\beqa
	\{\phi(\mb{x}),\pi(\mb{y})\} &=&  i_{\un{V}_{\phi(\mb{x})}} i_{\un{V}_{\pi(\mb{y})}}\Omega \nnb \\
	&=& \frac{1}{\sqrt{G(\mb{x})}}\delta^{(D)}(\mb{x}-\mb{y})\ .
\eeqa
This then tells us that the correct commutation relation for the operators $\hat{\phi}(\mb{x})$ and $\hat{\pi}(\mb{y})$ in the
quantized theory on curved space is
\beq
	[\hat{\phi}(\mb{x}),\hat{\pi}(\mb{y})] = \frac{i}{\sqrt{G(\mb{x})}}\delta^{(D)}(\mb{x}-\mb{y})\ .
\eeq

Given this form of $\Omega$, we can also work out a general formula for the Poisson bracket of any two functionals $F_1$
and $F_2$ of the phase space variables. To do this we need to first solve Eq.~\eqref{eq:vector-field}
for the vector field $\un{V}_F$ corresponding to a given functional $F$. If we write the vector field $\un{V}_F$ as
\beq
	\un{V}_F= \int d^D\mb{x}\ \left( V^{\phi}_F \frac{\delta}{\delta \phi(\mb{x})} + V^{\pi}_F \frac{\delta}{\delta \pi(\mb{x})} \right)\ ,
\eeq
then the solution of Eq.~\eqref{eq:vector-field} for the components of $\un{V}_F$ is 
\begin{subequations}
\beqa
	V^{\phi}_F &=& \frac{1}{\sqrt{G(\mb{x})}}\frac{\delta F}{\delta \pi(\mb{x})} \\
	V^{\pi}_F &=& -\frac{1}{\sqrt{G(\mb{x})}}\frac{\delta F}{\delta \phi(\mb{x})}\ .
\eeqa
\end{subequations}
Plugging into Eq.~\eqref{eq:PB}, we find that the Poisson bracket of any two functionals $F_1$ and $F_2$ in the theory of a 
free scalar field on a curved space is given by
\beq
	\{F_1,F_2\} = \int d^D\mb{x} \frac{1}{\sqrt{G(\mb{x})}}\left(\frac{\delta F_1}{\delta \phi(\mb{x})}\frac{\delta F_2}{\delta \pi(\mb{x})} - \frac{\delta F_1}{\delta \pi(\mb{x})}\frac{\delta F_2}{\delta \phi(\mb{x})}  \right)\ .
\eeq
The only modification from the usual Poisson bracket on flat space is the extra factor of $\frac{1}{\sqrt{G(\mb{x})}}$.

Now we combine this information with Dirac's procedure for dealing with constraints in the Hamiltonian formalism to derive
the commutation relations for the NLSM with theta term on the curved space $\mathcal{M}$. The action for the $O(D+2)$ NLSM 
with theta term on curved space is $S= \int d^{D+1}x\ \sqrt{G}\mathcal{L}$ with
\beq
	\mathcal{L}= \frac{1}{2f}(\pd^{\mu}n_a)(\pd_{\mu} n^a) + \frac{1}{\sqrt{G}}\frac{\theta}{\A_{D+1}}B_a(\pd_t n^a)\ ,
\label{eq:NLSM-Lag-curved}
\eeq
where the contraction of Greek (spacetime) indices is now done with the metric $g_{\mu\nu}$ from Eq.~\eqref{eq:metric}, and
$B_a$ was defined in Eq.~\eqref{eq:Ba}. The momentum conjugate to $n^a$ is now
\beq
	\pi_a = \frac{\pd \mathcal{L}}{\pd(\pd_t n^a)}=  \frac{1}{f}(\pd_t n_a) + \frac{1}{\sqrt{G}}\frac{\theta}{\A_{D+1}}B_a\ ,
\eeq
and the Hamiltonian on curved space takes the form
\begin{widetext}
\beq
	H= \int d^D\mb{x}\ \sqrt{G}\ \left\{ \frac{f}{2}\left(\pi_a - \frac{1}{\sqrt{G}}\frac{\theta}{\A_{D+1}}B_a\right)^2 + \frac{1}{2f}G^{ij}\pd_i n^a \pd_j n_a \right\}\ . \label{eq:NLSM-ham-curved}
\eeq
Finally, from our discussion above on the canonical formalism for a single scalar field on curved space, we know that the correct
Poisson bracket for two functionals $F_1$ and $F_2$ in the NLSM on curved space is
\beq
	\{F_1,F_2\}= \int d^D\mb{x}\ \frac{1}{\sqrt{G(\mb{x})}} \left( \frac{\delta F_1}{\delta n^a(\mb{x})}\frac{\delta F_2}{\delta \pi_a(\mb{x})}- \frac{\delta F_1}{\delta \pi_a(\mb{x})}\frac{\delta F_2}{\delta n^a(\mb{x})}  \right)\ . \label{eq:PB-curved}
\eeq
\end{widetext}

Using this Poisson bracket we may now proceed as in Sec.~\ref{sec:NLSM-quantization} and use Dirac's procedure for handling 
constraints to quantize the NLSM on curved space. We skip the details as they are very similar to those in 
Sec.~\ref{sec:NLSM-quantization}, and just present the results. The NLSM with theta term on curved space is again characterized
by two second class constraints,
\begin{subequations}
\beqa
	\psi_1(\mb{x}) &=& n^a(\mb{x})n_a(\mb{x})-1 \\
	\psi_2(\mb{x}) &=& n^a(\mb{x})\pi_a(\mb{x})\ .
\eeqa
\end{subequations}
The Poisson bracket of these constraints, computed using the Poisson bracket of Eq.~\eqref{eq:PB-curved} for the NLSM on
curved space, is $\{\psi_I(\mb{x}),\psi_J(\mb{y})\} = M_{IJ}(\mb{x},\mb{y})$ with 
\beq
	M_{IJ}(\mb{x},\mb{y}) = \frac{2i}{\sqrt{G(\mb{x})}} (\sigma^y)_{IJ}\ r^2(\mb{x})\delta^{(D)}(\mb{x}-\mb{y})\ ,
\eeq
where $r^2(\mb{x})= n_a(\mb{x}) n^a(\mb{x})$.
Its inverse, which is needed to compute the Dirac brackets for the NLSM on curved space, is
\beq
	M^{-1}_{IJ}(\mb{x},\mb{y})= -\frac{i}{2}\sqrt{G(\mb{x})} (\sigma^y)_{IJ} \frac{1}{r^2(\mb{x})}\delta^{(D)}(\mb{x}-\mb{y})\ . \label{eq:M-inv-curved}
\eeq
The components $M^{-1}_{IJ}(\mb{x},\mb{y})$ can now be used to construct the classical Dirac brackets for the 
NLSM on curved space. Then, to quantize the NLSM on curved space we replace all functions 
$f(\mb{x})$ with operators $\hat{f}(\mb{x})$ on the Hilbert space, and we obtain the quantum commutation relations for the NLSM on 
curved space by replacing the Dirac brackets with commutators as in Sec.~\ref{sec:NLSM-quantization} for the NLSM on flat space.
In addition, as in Sec.~\ref{sec:NLSM-quantization}, we set the operator $\hat{r}^2(\mb{x})=1$, which is consistent since this operator 
commutes with all other operators in the Hilbert space. Therefore we find that the commutation relations for the NLSM with theta term on 
curved space are
\begin{widetext}
\begin{subequations}
\label{eq:comm-rels-NLSM-curved}
\beqa
	\left[\hat{n}^a(\mb{x}), \hat{n}^b(\mb{y})\right] &=& 0  \\
	\left[\hat{n}^a(\mb{x}), \hat{\pi}_b(\mb{y})\right] &=& \frac{i}{\sqrt{G(\mb{x})}}({\delta^a}_b - \hat{n}^a(\mb{x}) \hat{n}_b(\mb{y}))\delta^{(D)}(\mb{x}-\mb{y}) \\
	\left[\hat{\pi}_a(\mb{x}), \hat{\pi}_b(\mb{y})\right] &=& \frac{i}{\sqrt{G(\mb{x})}}(\hat{\pi}_a(\mb{x}) \hat{n}_b(\mb{y}) - \hat{\pi}_b(\mb{y}) \hat{n}_a(\mb{x}))\delta^{(D)}(\mb{x}-\mb{y})\ . 
\eeqa
\end{subequations}
\end{widetext}
Again, the only modification from the case of flat space is the extra factor of $\frac{1}{\sqrt{G(\mb{x})}}$. As in the
case on flat space, these commutation
relations also admit a functional Schroedinger representation in which $\hat{n}^a(\mb{x})$ acts as multiplication by $n^a(\mb{x})$
and now $\hat{\pi}_a(\mb{x})$ acts as the functional derivative operator
\beq
	\hat{\pi}_a(\mb{x})= -\frac{i}{\sqrt{G(\mb{x})}} \left({\delta_a}^b - n_a(\mb{x})n^b(\mb{x})\right)\frac{\delta}{\delta n^b(\mb{x})}\ . \label{eq:schro-rep-curved}
\eeq
In the next subsection we use this Schroedinger representation to solve for the ground state wave functional of the 
$O(D+2)$ NLSM in the disordered ($f \to\infty$) phase at $\theta=2\pi k$, $k\in\mathbb{Z}$.

\subsection{Ground state wave functional at large $f$ and $\theta= 2\pi k$, $k\in\mathbb{Z}$}

In the large $f$ limit the Hamiltonian operator for the $O(D+2)$ NLSM with theta term on the curved space
$\mathcal{M}$ takes the form
\beq
	\hat{H}= \int d^D\mb{x}\ \sqrt{G}\ \frac{f}{2}\left(\hat{\pi}_a - \frac{1}{\sqrt{G}}\frac{\theta}{\A_{D+1}}\hat{B}_a\right)^2\ , 
\eeq
where we again dropped the potential energy term with coefficient proportional to $\frac{1}{f}$. We now investigate the
form of the ground state wave functional of this theory in the case where $\theta=2\pi k$, $k\in\mathbb{Z}$, which is the
case where the NLSM is expected to describe an SPT phase on the curved space $\mathcal{M}$.
As in the case on flat space, the ground state wave functional is determined by the condition that it be annihilated by the operators 
\beq
	\hat{\mathcal{D}}^{(\theta)}_a= \hat{\pi}_a - \frac{1}{\sqrt{G}}\frac{\theta}{\A_{D+1}}\hat{B}_a\ ,\ a= 1,\dots,D+2\ . 
\label{eq:D-operator-curved-space}
\eeq
In the functional Schroedinger representation used in this paper this operator takes the form
\begin{widetext}
\beq
	\hat{\mathcal{D}}^{(\theta)}_a(\mb{x}) = \frac{1}{\sqrt{G(\mb{x})}} \left\{ -i\left({\delta_a}^b - n_a(\mb{x})n^b(\mb{x})\right)\frac{\delta}{\delta n^b(\mb{x})} - \frac{\theta}{\A_{D+1}}B_a(\mb{x}) \right\}\ .
\eeq
\end{widetext}
We see that the dependence of this operator on the metric of space is \emph{only} through the overall factor of 
$\frac{1}{\sqrt{G(\mb{x})}}$. This a consequence of the fact that the Dirac brackets for the NLSM on curved
space have an explicit dependence on $\sqrt{G(\mb{x})}$, while the theta term in the NLSM action is independent of the metric. 
This property of $\hat{\mathcal{D}}^{(\theta)}_a(\mb{x})$ is very important. It implies that the ground state wave functional 
at large $f$ and $\theta=2\pi k$ on the curved space $\mathcal{M}$ is \emph{independent} of the metric on $\mathcal{M}$, and
can be constructed in the exact same way as on flat space, i.e., the ground state wave functional is the exponential of a WZ term 
for the NLSM field, $\Psi_{\theta=2\pi k}[\mb{n}]= e^{-ik S_{WZ}[\mb{n}]}$. As we mentioned at the beginning of this 
section, for certain choices of $\mathcal{M}$ the standard construction of the WZ term fails, and so 
alternative constructions are needed. We now turn to a discussion of this issue.

The crucial property of the WZ term $S_{WZ}[\mb{n}]$, which allows for the construction of a functional annihilated by 
$\hat{\mathcal{D}}^{(\theta=2\pi k)}_a$, is the formula Eq.~\eqref{eq:WZ-variation} for its variation with respect to 
$n^a(\mb{x})$. We now review two different methods for constructing an action whose variation is given by
Eq.~\eqref{eq:WZ-variation}, and then we discuss specific examples of manifolds $\mathcal{M}$ where both constructions
fail. In these cases a third construction of the WZ term is available using the methods outlined in 
Ref.~\cite{alvarez1985topological}. In Appendix~\ref{app:WZ-intrinsic} we give an explicit example of the construction of the
WZ term using the methods of Ref.~\cite{alvarez1985topological} in the simple case where the dimension of the space
$\mathcal{M}$ is $D=1$.

The first construction of the WZ term that we discuss is the standard construction that appears in the 
literature~\cite{witten1983global}. This construction uses a higher-dimensional manifold $\mathcal{B}$ which has
$\mathcal{M}$ as its boundary. In Sec.~\ref{sec:wavefcn-gap} we discussed this construction on flat space 
$\mathcal{M}=\mathbb{R}^D$, and we now discuss how it works for a general curved spatial manifold $\mathcal{M}$.  
For the standard construction of the WZ term for the NLSM field we first look for a $(D+1)$-dimensional manifold $\mathcal{B}$ which 
has $\mathcal{M}$ as its boundary, $\pd \mathcal{B}=\mathcal{M}$. Then, for a given NLSM field configuration $n^a$ on 
$\mathcal{M}$, we construct an extension $\tilde{n}^a$ of the NLSM field configuration into the bulk of the manifold $\mathcal{B}$ 
such that $\tilde{n}^a|_{\pd\mathcal{B}}= n^a$. Finally, using the extended manifold $\mathcal{B}$ and the extension 
$\tilde{n}^a$ of the NLSM field, the standard construction of the WZ term for $\mb{n}$ is given by
\beq
	S_{WZ}[\mb{n}] = \frac{2\pi}{\A_{D+1}}\int_{\mathcal{B}}\tilde{\mb{n}}^*\omega_{D+1}\ , \label{eq:WZ-term-curved}
\eeq
where $\tilde{\mb{n}}^*\omega_{D+1}$ denotes the pullback of the volume form $\omega_{D+1}$ of $S^{D+1}$ to the
extended space $\mathcal{B}$ via the map $\tilde{\mb{n}}: \mathcal{B}\to S^{D+1}$. Since in this construction the WZ term 
depends on the choice of the manifold $\mathcal{B},$ and the choice of extension of the NLSM field $\tilde{\mb{n}}$, we need
to check that the exponential $e^{-ik S_{WZ}[\mb{n}]}$ is independent of these choices in order for the wave functional to 
be well-defined. 

The exponential of the WZ term constructed in this way will be well-defined if it is independent of the specific choices of 
extended manifold $\mathcal{B}$ and field extension $\tilde{n}^a$. 
To check this, suppose we have two different choices of extended manifold
$\mathcal{B}$ and $\mathcal{B}'$, with $\pd\mathcal{B}=\pd\mathcal{B}'=\mathcal{M}$, and two different field extensions 
$\tilde{\mb{n}}$ and $\tilde{\mb{n}}'$, with 
$\tilde{\mb{n}}|_{\pd\mathcal{B}}= \tilde{\mb{n}}'|_{\pd\mathcal{B}'}=\mb{n}$. Let $S_{WZ}[\mb{n}]$ and $S'_{WZ}[\mb{n}]$
be the WZ terms defined using $(\mathcal{B},\tilde{\mb{n}})$ and $(\mathcal{B}',\tilde{\mb{n}}')$, respectively. Then we 
can write
\beq
	e^{-ik S_{WZ}[\mb{n}]}= e^{-ik\left(S_{WZ}[\mb{n}]-S'_{WZ}[\mb{n}] \right)}e^{-ik S'_{WZ}[\mb{n}]}\ .
\eeq
If follows from this expression that the exponential of the WZ term will be well-defined if the difference
$S_{WZ}[\mb{n}]-S'_{WZ}[\mb{n}]$ of the two WZ terms is an integer multiple of $2\pi$ (we assume $k\in\mathbb{Z}$), since
in that case we have $e^{-ik S_{WZ}[\mb{n}]}= e^{-ik S'_{WZ}[\mb{n}]}$.
The difference of WZ terms is in turn equivalent to a single integral 
\beq
	I[\tilde{\mb{n}}'']= \frac{2\pi}{\A_{D+1}}\int_{\mathcal{X}}\tilde{\mb{n}}''^*\omega_{D+1}\ , \label{eq:integral-to-check}
\eeq
where $\mathcal{X}$ is a closed $(D+1)$-dimensional manifold formed by gluing $\mathcal{B}$ to $\mathcal{B}'$ along their
common boundary $\mathcal{M}$, and where $\tilde{\mb{n}}''$ is the NLSM field configuration on the entire $(D+1)$-dimensional 
manifold $\mathcal{X}$ formed in this way ($\tilde{\mb{n}}$ and $\tilde{\mb{n}}'$ agree at the boundary where the gluing takes
place, and on the rest of $\mathcal{X}$ they define the configuration $\tilde{\mb{n}}''$). 
Since we are dealing with orientable manifolds, we must specify the orientation
of the boundaries of $\mathcal{B}$ and $\mathcal{B}'$ when we glue them together to construct $\mathcal{X}$. In the 
construction of $\mathcal{X}$ discussed here the manifolds $\mathcal{B}$ and $\mathcal{B}'$ are glued together in a such a way that the 
orientation of $\pd\mathcal{B}'$ is opposite to the orientation of $\pd\mathcal{B}$. This choice of orientations is forced on us
because we are considering the \emph{difference} of WZ terms.


We see that in order to determine whether the exponential of the WZ term is well-defined, it suffices to check that the integral in 
Eq.~\eqref{eq:integral-to-check} is an integer multiple of $2\pi$ for \emph{any} closed $(D+1)$-dimensional manifold 
$\mathcal{X}$ and $\emph{any}$ NLSM field configuration $\tilde{\mb{n}}''$ on $\mathcal{X}$. To see that this is indeed the
case, we note that 
\beq
	\int_{\mathcal{X}}\tilde{\mb{n}}''^*\omega_{D+1} = \text{deg}[\tilde{\mb{n}}'']\int_{S^{D+1}}\omega_{D+1}\ ,
\eeq
where $\text{deg}[\tilde{\mb{n}}''] \in \mathbb{Z}$ is the \emph{degree} of the map $\tilde{\mb{n}}'': \mathcal{X}\to S^{D+1}$. 
It is an integer which counts how many times the space $\mathcal{X}$ is ``wrapped" around the sphere $S^{D+1}$ by the
map $\tilde{\mb{n}}''$ (see, for example, Ch. VI of Ref.~\cite{flanders1963differential}). Combining this with the fact that 
$\int_{S^{D+1}}\omega_{D+1}= \A_{D+1}$, we find that
\beq
	I[\tilde{\mb{n}}'']= 2\pi\ \text{deg}[\tilde{\mb{n}}'']\ ,
\eeq
which proves that the exponential $e^{-i k S_{WZ}[\mb{n}]}$ of the WZ term is well-defined for integer 
$k$.


Besides the standard construction of the WZ term using the higher-dimensional manifold $\mathcal{B}$,
it is also possible to define $S_{WZ}[\mb{n}]$ as a functional integral in a theory of fermions defined on the manifold $\mathcal{M}$. 
This construction relies on a result of Abanov and Wiegmann, who constructed theories of fermions coupled
to an NLSM field $\mb{n}$ which produce a WZ term for $\mb{n}$ after integrating out the fermions~\cite{AbanovWiegmann}. 
The coupling of the fermions to the NLSM field involves a mass parameter $M$, and the partition function
$Z[\mb{n}]$ for the theory of fermions coupled to $\mb{n}$ can be computed in a gradient expansion in powers of 
$\frac{1}{M}$. From this partition function one can define an effective action for the NLSM field via 
$S_{eff}[\mb{n}]= -\ln(Z[\mb{n}])$. In Ref.~\cite{AbanovWiegmann} the authors calculated
the variation of $S_{eff}[\mb{n}]$ with respect to $n^a(\mb{x})$, and they showed that
the imaginary part of this variation has exactly the form of Eq.~\eqref{eq:WZ-variation}. Therefore, the results of 
Ref.~\cite{AbanovWiegmann} imply that one can \emph{define} the WZ term using the partition function $Z[\mb{n}]$ as 
\beq
	S_{WZ}[\mb{n}]= -\mathfrak{Im}[\ln(Z[\mb{n}])]\ ,
\eeq
where $\mathfrak{Im}[\cdots]$ denotes the imaginary part. We also note that this definition naturally produces a WZ term with an integer 
level $k= \text{sgn}[M]N_F$, where $N_F$ (a positive integer) is the number of flavors of fermions that couple to $\mb{n}$.
In particular, it does not seem possible to generate a WZ term with fractional level in this way.

So far we have presented two different ways of constructing the WZ term for the NLSM field $\mb{n}$ on
a curved space $\mathcal{M}$. One interesting aspect of considering the NLSM on general spaces $\mathcal{M}$
is that there are certain choices of $\mathcal{M}$ where neither of these constructions works. This can be seen as follows.
First, the standard construction of the WZ term requires that there exists a $\mathcal{B}$ such that 
$\pd\mathcal{B}=\mathcal{M}$. However, there are some manifolds $\mathcal{M}$ which cannot be realized as the
boundary of \emph{any} higher-dimensional manifold. The precise conditions for $\mathcal{M}$ to be a boundary are given
by the following theorem  (see, for example, Ch. 4 of Ref.~\cite{milnor}).

\textbf{Theorem} (Thom)\textbf{:} If all of the \emph{Stiefel-Whitney numbers} of $\mathcal{M}$ are zero, then $\mathcal{M}$ can be 
realized as the boundary of some smooth compact manifold $\mathcal{B}$.

In dimensions $D=1,2,$ and $3$, every orientable $\mathcal{M}$ is a boundary. The situation becomes more interesting
for $D\geq 4$. In the case that $D\equiv 0$ mod $4$, it is easy to 
construct simple examples of orientable manifolds $\mathcal{M}$ which are not a boundary by taking products of $C\mathbb{P}^{2r}$ for 
positive integer $r$, for example $C\mathbb{P}^2$ in $D=4$ and $C\mathbb{P}^2\times C\mathbb{P}^2$ and $C\mathbb{P}^4$ 
in $D=8$. Orientable manifolds which are not a boundary also exist in dimensions $D\geq 4$ where $D$ is not equivalent to zero modulo
four, for example in $D=5,9,10$, and $11$~\cite{milnor}. Thus, we find that for many values of 
$D>3$ there are choices of $\mathcal{M}$ where the standard construction of the WZ term fails. 

The second construction of the WZ term, defined using a path integral for fermions on $\mathcal{M}$, can fail if the
manifold $\mathcal{M}$ does not admit a spin structure. If $\mathcal{M}$ does not admit a spin structure then it is not possible to formulate
a consistent theory of fermions on $\mathcal{M}$.
The technical requirement for the existence of a spin structure on 
$\mathcal{M}$ is that $w_2 \in H^2(\mathcal{M},\mathbb{Z}_2)$, the second \emph{Stiefel-Whitney class} of $\mathcal{M}$, must 
vanish~\cite{EGH}. Note that we assume that $\mathcal{M}$ is orientable, and so we also require that the first 
Stiefel-Whitney class of $\mathcal{M}$, $w_1 \in H^1(\mathcal{M},\mathbb{Z}_2)$, is trivial. In fact, a spin structure cannot
be defined on an unorientable manifold, so this condition is crucial for the second construction of the WZ term using a path integral
over fermions. 

In some cases the first construction can fail but the second construction works. One example of such a case can be found in 
$D=4$ when $\mathcal{M}$ is taken to be the \emph{Kummer surface}. This four-dimensional manifold is not a boundary but does
admit a spin structure (see, for example, Ch. XI of Ref.~\cite{kirby2006topology}). 
A particularly interesting example, also in $D=4$, is the choice $\mathcal{M}=C\mathbb{P}^2$. In this
case \emph{both} constructions fail. Therefore we find that in general a third construction of the WZ term is needed. This 
third construction should not require $\mathcal{M}$ to be a boundary, and it should also not require that $\mathcal{M}$ admit
a spin structure. We refer to such a construction as an ``intrinsic construction" since it does not require an extension $\mathcal{B}$ 
of $\mathcal{M}$. It turns out that such a construction does exist. In particular, in Ref.~\cite{alvarez1985topological}
Alvarez explained how to carry out this construction in detail using the language of \v{C}ech cohomology. In 
Appendix~\ref{app:WZ-intrinsic} we give an example of this type of construction in the simple case that $D=1$.

To summarize, we find that the ground state wave functional of the $O(D+2)$ NLSM at $\theta=2\pi k$, $k\in\mathbb{Z}$, in the disordered 
($f\to\infty$) phase is
\beq
	\Psi_{\theta=2\pi k}[\mb{n}]= e^{-ikS_{WZ}[\mb{n}]}\ ,
\eeq
where $S_{WZ}[\mb{n}]$ is a suitably defined WZ term for the NLSM field $\mb{n}$. 
As we discussed above, the specific construction of the WZ term will depend on the particular spatial manifold $\mathcal{M}$, 
but the WZ term always exists. The ground state wave functional has several important properties. First, it is independent of the
metric of space, which shows that the disordered phase of the NLSM at $\theta=2\pi k$ is a topological phase. Second, it 
possesses the full $SO(D+2)$ symmetry of the action for the NLSM with theta term. These two properties taken together
imply that $\Psi_{\theta=2\pi k}[\mb{n}]$ can be the ground state wave functional of an SPT phase.
As in Sec.~\ref{sec:wavefcn-gap}, we can also understand the ground state wave functional as arising from a unitary 
transformation by the operator $\hat{\mathcal{U}}^{(k)}$ from Eq.~\eqref{eq:unitary-operator}, which transforms the 
Hamiltonian at $\theta=2\pi k$ into the Hamiltonian at $\theta=0$ as in Eq.~\eqref{eq:unitary-trans}. The ground state of the 
theory at $\theta=2\pi k$ can then be obtained by applying $\hat{\mathcal{U}}^{(k)}$ to the ground state at $\theta=0$ as in
Eq.~\eqref{eq:states-0-2pik}.

\subsection{Uniqueness of the ground state and the energy gap at large $f$ and $\theta=2\pi k$, $k\in\mathbb{Z}$}

In this section we study the spectrum of the NLSM on the curved space $\mathcal{M}$ by using
a \emph{triangulation} of the manifold to implement a lattice-like regularization of the NLSM Hamiltonian 
on $\mathcal{M}$. Using this regularization we demonstrate the uniqueness of the ground state and the existence of an
energy gap in the $O(D+2)$ NLSM in the disordered ($f\to\infty$) phase at $\theta=0$ on $\mathcal{M}$.
Since the NLSM Hamiltonian at $\theta=2\pi k$, $k\in\mathbb{Z}$, is related to the Hamiltonian at $\theta=0$ by a 
unitary transformation, it will follow from the results of this section that the ground state of the NLSM in the disordered phase at
$\theta=2\pi k$, $k\in\mathbb{Z}$, on $\mathcal{M}$ is also unique for all $k.$ Thus, this subsection completes 
our proof of the absence of topological order in the $O(D+2)$ NLSM in the disordered phase at $\theta=2\pi k$, $k\in\mathbb{Z}$.

We start by recalling a few basic facts about triangulations of smooth manifolds, following the discussion in sections 3.2 and
5.3.2 of Ref.~\cite{prasolov2006elements}. Intuitively, a triangulation of a manifold is an approximation of 
the manifold by generalized triangles called \emph{simplices}. A $0$-simplex is a point, a $1$-simplex is a line segment, a 
$2$-simplex is a triangle, a $3$-simplex is a tetrahedron, and so on\footnote{The \emph{standard} $n$-simplex is the 
region $\Delta^n \subset \mathbb{R}^{n+1}$ defined by 
$\Delta^n= \left\{\sum_{i=1}^{n+1} y^i = 1\ ,\ y^i \geq 0 \ \forall\ i\right\}$.}. A simplicial complex $K$ is a set of 
simplices in $\mathbb{R}^n$ such that (i) all faces of simplices from $K$ belong to $K$, (ii) the intersection of any two simplices from
$K$ is a face for each of them, and (iii) any point that belongs to one of the simplices from $K$ has a neighborhood which 
intersects only finitely many simplices from $K$. For any such simplicial complex $K$, the space $|K|$ is the topological space which is 
the union of all simplices of $K$ with the topology induced by $\mathbb{R}^n$. Finally, a \emph{triangulation} of a manifold 
$\mathcal{M}$ is a homeomorphism $\rho:\mathcal{M} \to |K|$, where $K$ is a simplicial complex in $\mathbb{R}^n$ for some $n$ 
(with $n$ greater than or equal to the dimension of $\mathcal{M}$). Any smooth closed manifold $\mathcal{M}$ admits a 
triangulation.

Now let us pick particular triangulation $(\rho,|K|)$ of $\mathcal{M}$. Let $\al=1,\dots,N_D$, label the distinct $D$-simplices
in $|K|$, and define $\mathcal{S}_{\al} \subset \mathcal{M}$ to be the inverse image of the $D$-simplex $\al$ under the map
$\rho$ ($\rho$ is a homeomorphism so it is invertible). 
We will also refer to $\mathcal{S}_{\al}$ as a $D$-simplex. For our purposes, the key property of the triangulation is that 
it allows for a decomposition of $\mathcal{M}$ as $\mathcal{M}= \sum_{\al=1}^{N_D}\mathcal{S}_{\al}$, 
where the sum is the composition of oriented $D$-chains.
To set up a ``lattice" on $\mathcal{M}$, we can then pick an arbitrary point $p_{\al}$ in each 
$\mathcal{S}_{\al}$ to be the points of the lattice. In the lattice regularization on flat space each lattice point was associated with a 
hypercubic unit cell of volume $a^D$, where $a$ was the lattice spacing. In our regularization on curved space each point 
$p_{\al}$ is associated with a $D$-simplex $\mathcal{S}_{\al}$, and each such $D$-simplex has a volume given by 
\beq
	\text{Vol}_{\al}= \int_{\mathcal{S}_{\al}}\text{Vol}_{\mathcal{M}}\ ,
\eeq
where $\text{Vol}_{\mathcal{M}}$ is the volume form on $\mathcal{M}$ determined by its Riemannian metric. 
In a system of local coordinates $(U,\phi)$, $U\subset \mathcal{M}$, $\phi: U \to \mathbb{R}^D$, one has 
$\text{Vol}_{\mathcal{M}} = d^D\mb{x}\ \sqrt{G(\mb{x})}$, if $\mb{x} \in\mathbb{R}^D$ denotes the image of a point
$p\in U$ under $\phi$.

Using this regularization, integration of a function $f(p)$ over $\mathcal{M}$, weighted by the volume form 
$\text{Vol}_{\mathcal{M}}$, can be discretized as 
\beq
	\int_{p\in\mathcal{M}}\text{Vol}_{\mathcal{M}}\ f(p) \to \sum_{\al=1}^{N_D} \text{Vol}_{\al}\ f(p_{\al})\ . 
\eeq
In addition, we can define a Dirac delta function $\delta(p,p')$ on $\mathcal{M}$ by
\beq
	\int_{p\in\mathcal{M}}\text{Vol}_{\mathcal{M}}\ f(p)\ \delta(p,p') = f(p')\ .
\eeq
Again, in a system of local coordinates $(U,\phi)$ 
one has $\delta(p,p') = \frac{1}{\sqrt{G(\mb{x})}}\delta^{(D)}(\mb{x}-\mb{x}')$, where 
$\mb{x},\mb{x}'\in\mathbb{R}^D$ are the images of $p,p'$ under the map $\phi$. Then $\delta(p,p')$ is discretized as
\beq
	\delta(p,p') \to \frac{1}{\text{Vol}_{\al}}\delta_{\al\beta}\ 
\eeq
if $p=p_{\al}$ and $p'=p_{\beta},$ and $\delta_{\al\beta}$ is the ordinary Kronecker delta.

Throughout this subsection we considered the NLSM in a particular coordinate patch of $\mathcal{M}$ with local coordinates
$\mb{x}$. However, using the Dirac delta function $\delta(p,p')$ on $\mathcal{M}$ we can also write the commutation relations
for the NLSM on $\mathcal{M}$ as
\begin{subequations}
\label{eq:comm-rels-NLSM-curved-no-coords}
\begin{align}
	\left[\hat{n}^a(p), \hat{n}^b(p')\right] &= 0  \\
	\left[\hat{n}^a(p), \hat{\pi}_b(p')\right] &= i({\delta^a}_b - \hat{n}^a(p) \hat{n}_b(p'))\delta(p,p') \\
	\left[\hat{\pi}_a(p), \hat{\pi}_b(p')\right] &= i(\hat{\pi}_a(p) \hat{n}_b(p') - \hat{\pi}_b(p') \hat{n}_a(p))\delta(p,p')\ . 
\end{align}
\end{subequations}
For the lattice regularization of the NLSM using the triangulation discussed above, we define lattice variables by
\bseq
\beqa
	\hat{n}^a(p_{\al}) &=& \hat{n}^a_{\al} \\
	\hat{\pi}_a(p_{\al}) &=& \frac{1}{\text{Vol}_{\al}}\hat{\pi}_{a,\al} 
\eeqa
\eseq
for all points $p_{\al}$ in the lattice constructed from the triangulation. The lattice variables obey the commutation relations
\begin{subequations}
\beqa
	\left[\hat{n}^a_{\al}, \hat{n}^b_{\beta}\right] &=& 0  \\
	\left[\hat{n}^a_{\al}, \hat{\pi}_{b,\beta}\right] &=& i({\delta^a}_b - \hat{n}^a_{\al} \hat{n}_{b,\beta})\delta_{\al\beta} \\
	\left[\hat{\pi}_{a,\al}, \hat{\pi}_{b,\beta}\right] &=& i(\hat{\pi}_{a,\al} \hat{n}_{b,\beta} - \hat{\pi}_{b,\beta} \hat{n}_{a,\al})\delta_{\al\beta}\  ,
\eeqa
\end{subequations}
and the regularized Hamiltonian takes the form
\beq
	\hat{H}(\rho,|K|)= \frac{f}{2}\sum_{\al=1}^{N_D}\frac{1}{\text{Vol}_{\al}}\hat{\pi}_{a,\al} {\hat{\pi}^a}_{\al} \ . \label{eq:Ham-reg-curved}
\eeq
Here we have written $\hat{H}(\rho,|K|)$ to indicate the dependence of the regularized Hamiltonian on the choice of a triangulation 
$(\rho,|K|)$ of $\mathcal{M}$.

Just as in the case on flat space in Sec.~\ref{sec:wavefcn-gap}, the regularized Hamiltonian breaks up into a sum of decoupled 
Hamiltonians for an $O(D+2)$ NLSM in one spacetime dimension. Therefore we can again apply the results of 
Appendix~\ref{app:NLSM-one-dim} to deduce that the ground state of the $O(D+2)$ NLSM in the 
disordered ($f\to\infty$) phase on $\mathcal{M}$ is 
\emph{unique}, and corresponds to the state where $\hat{\pi}_{a,\al} {\hat{\pi}^a}_{\al} = 0$ $\forall\ \al$. Unlike the case
of flat space, however, there is now a different energy cost to creating an excitation on the different sites $p_{\al}$ in the lattice.
This is due the presence of the factor $(\text{Vol}_{\al})^{-1}$ in the summand in Eq.~\eqref{eq:Ham-reg-curved} 
(compare to Eq.~\eqref{eq:reg-Ham} on flat space, which just contains an overall factor of $a^{-D}$). 
It follows that the energy gap for the regularized $O(D+2)$ NLSM on $\mathcal{M}$ is given by
\beq
	m(\rho,|K|) = \frac{f}{2}\frac{(D+1)}{\overline{\text{Vol}}(\rho,|K|)}\ ,
\eeq
where we defined $\overline{\text{Vol}}(\rho,|K|)$ to be the volume of the \emph{largest} $D$-chain $\mathcal{S}_{\al}$
in the triangulation $(\rho,|K|)$ of $\mathcal{M}$,
\beq
	\overline{\text{Vol}}(\rho,|K|)= \text{max}\big\{ \text{Vol}_{\al} | \al \in \{1,\dots,N_D\} \big\}\ .
\eeq

Thus, we have proven the uniqueness of the ground state and the existence of an energy gap in the $O(D+2)$ NLSM
in the disordered phase at $\theta=0$. From the previous subsection we know that the Hamiltonian for the $O(D+2)$ NLSM
at $\theta=2\pi k$, $k\in\mathbb{Z}$, is related to the Hamiltonian at $\theta=0$ by a unitary transformation, so the
result in this section then implies uniqueness of the ground state and the existence of an energy gap in the $O(D+2)$ NLSM
in the disordered phase at $\theta=2\pi k$, $k\in\mathbb{Z}$, for all $k$. This result completes the demonstration that
the $O(D+2)$ NLSM in the disordered phase at $\theta=2\pi k$ is a suitable model for SPT phases of bosons.


\section{Conclusion}
\label{sec:conclusion}

In this paper we performed an explicit study of the $O(D+2)$ NLSM with theta term, in its disordered phase and
with theta angle $\theta=2\pi k$, $k\in\mathbb{Z}$, on arbitrary smooth, closed, connected, oriented $D$-dimensional 
spatial manifolds $\mathcal{M}$. We showed that in this parameter regime the ground state of the NLSM on $\mathcal{M}$ is 
unique, and there is a finite energy gap to the lowest lying excited states. In addition, we showed that the ground state wave
functional of the NLSM on $\mathcal{M}$ is independent of the metric on $\mathcal{M}$, and takes the form of an exponential
of a WZ term for the NLSM field $\mb{n}$, just like in the case on flat space~\cite{xu2013wave}. These results taken together
imply that the $O(D+2)$ NLSM, in the disordered phase with $\theta=2\pi k$, $k\in\mathbb{Z}$, is a suitable model for an
SPT phase of bosons. In particular, our results show that this model \emph{does not} possess topological order. Thus, our work 
places the NLSM approach to SPT phases of Ref.~\cite{CenkeClass1} on solid ground. 

We close the paper with some additional comments and suggestions for future work. First, we mention one puzzle associated
with the NLSM description of SPT phases. In several recent
works~\cite{kapustin2014symmetry,kapustin2015fermionic,hsieh2016global,witten2015fermion}
 it was shown that important information about 
the classification of SPT phases with time-reversal symmetry $\mathbb{Z}^T_2$ can be extracted from the partition functions of 
these phases on unorientable Euclidean spacetime manifolds. 
For example, in two spacetime dimensions there is a single nontrivial bosonic SPT phase protected only by $\mathbb{Z}^T_2 $, 
and in Ref.~\cite{kapustin2014symmetry} it was shown that this SPT phase can be detected by its partition function 
$Z_{R\mathbb{P}^2}=-1$ on the spacetime $R\mathbb{P}^2$. An NLSM description of this SPT phase is available (see Sec. IV.B of
Ref.~\cite{CenkeClass1}), but it seems problematic to calculate the partition function of the NLSM on 
$R\mathbb{P}^2$. Mathematically, the issue is that the theta term for the $O(D+2)$ NLSM involves the pullback to spacetime of the 
volume form on $S^{D+1}$, and this pullback does not seem to make sense when the spacetime is not orientable. Therefore
it would be interesting to see if there is some way to make sense of the NLSM description of SPT phases on unorientable 
spacetime manifolds. 

A second possible direction for future work would be to extend the analysis of this paper to the case of the $O(D+2)$ NLSM in the 
disordered limit when the theta angle $\theta$ is an \emph{odd} multiple of $\pi$. In Ref.~\cite{xu2013nonperturbative} the authors
used a qualitative argument to map out the phase diagram of the $O(4)$ NLSM in $D=2$ spatial dimensions with theta term and
coefficient $\theta=\pi$. They proposed two possible phases for this theory when the coupling constant $f$ is large: (i) a gapless phase 
realized at some finite but large value of $f$, and (ii) a gapped phase which is realized in
the extreme disordered limit of $f\to \infty$. In addition, the authors of Ref.~\cite{xu2013nonperturbative} argued that the ground 
state in the gapped phase should be doubly degenerate (see also Ref.~\cite{SenthilFisher} on this point). It would be interesting to 
investigate the $f\to\infty$ phase directly within the canonical formalism, with the goal of proving that in this limit the spectrum
is indeed gapped and that the ground state is doubly degenerate. It would also be interesting to investigate the dependence of the ground 
state degeneracy on the topology of the spatial manifold.
This problem is quite interesting for the following reason. Typically, the boundary theory of an SPT phase is expected to 
preserve the symmetry of the SPT phase and be gapless, or to spontaneously break the symmetry in some way (which may
lead to a gapped boundary theory). However, at
the $(2+1)$-dimensional boundary of a $(3+1)$-dimensional SPT phase (and presumably also in higher dimensions) 
there is a third possibility: the boundary theory can be gapped
and symmetric, but it must also possess intrinsic topological order~\cite{VS2013,metlitski2013bosonic}. 
It is likely that the $O(4)$ NLSM, in the disordered phase and at $\theta=\pi$, can describe such a gapped, symmetry-preserving, and 
topologically-ordered surface state of the bosonic topological insulator phase in $3+1$ dimensions~\cite{VS2013}. Therefore it would be 
interesting to give a proof that the $O(4)$ NLSM in this parameter regime really does possess intrinsic topological order.
As we discussed in Sec.~\ref{sec:wavefcn-gap}, the unitary transformation which removes the 
theta angle from the Hamiltonian can only be performed when $\theta$ is a multiple of $2\pi$, which means that completely new 
methods will be needed to solve the problem of the NLSM in the disordered phase at $\theta=\pi$.

\acknowledgements

We thank A. Kapustin and M. Stone for helpful discussions on the content of this paper. 
MFL and TLH acknowledge support from the ONR YIP Award N00014-15-1-2383.  We also gratefully acknowledge the support of the Institute 
for Condensed Matter Theory at the University of Illinois at Urbana-Champaign.

\appendix

\section{Canonical quantization of the $O(N)$ NLSM in $0+1$ dimensions}
\label{app:NLSM-one-dim}

In this appendix we review the solution of the $O(N)$ NLSM in one spacetime dimension using the commutation relations
of Eqs.~\eqref{eq:comm-rels-NLSM} and the Schroedinger representation in Eq.~\eqref{eq:schro-rep}. 
We use this solution in Secs.~\ref{sec:wavefcn-gap} and
\ref{sec:NLSM-quantization-curved} to compute the energy gap in regularized versions of the $O(D+2)$ NLSM 
in the disordered ($f\to\infty$) limit and with $\theta=2\pi k$, $k\in\mathbb{Z}$, on flat and curved space, respectively. 
The $O(N)$ NLSM in one spacetime dimension is equivalent to the quantum mechanics problem of a free particle
in $N$ spatial dimensions but confined to the surface of the unit sphere $S^{N-1}$. This is a famous problem in the
quantization of constrained systems and has been studied by many 
authors~\cite{podolsky1928quantum,dewitt1957dynamical,kleinert1997proper,neto1999does,neves2000stuckelberg,
hong2000improved,abdalla2001quantisation,scardicchio2002classical,hong2004gauged}. 
One finds that the quantum mechanical Hamiltonian is proportional to the Laplace-Beltrami operator $\Delta^{(N-1)}_{LB}$ on the 
sphere $S^{N-1}$, so that the energy spectrum is given in terms of the eigenvalues of $\Delta^{(N-1)}_{LB}$. 
However, there is some controversy in the literature about whether an additional constant term, depending
only on $N$, should appear in the quantum Hamiltonian for this problem. This constant term is irrelevant for the application to our discussion 
in the context of Secs.~\ref{sec:wavefcn-gap} 
and \ref{sec:NLSM-quantization-curved}, in which we are interested only in the \emph{difference} between the energy of the
ground state and first excited state. Therefore in this appendix we give a straightforward analysis of the 
$O(N)$ NLSM in one spacetime dimension, without worrying about subtleties (e.g., Weyl-ordering of operators to define the
quantum Hamiltonian~\cite{kleinert1997proper,neto1999does,hong2000improved}) which could lead to an extra constant shift in the 
energy spectrum. Readers interested in the subtleties associated with this constant term should consult the references cited in this 
paragraph. 

\subsection{Hamiltonian, commutation relations, and the energy spectrum}

We consider the $O(N)$ NLSM in one spacetime dimension. For general $N$ this system does not admit a 
theta term (a theta term is possible at $N=2= D+2$ since $D=0$ here). However, in this appendix we are only interested in 
discussing the quantization of the theory without theta term, and so we consider the case of a general $N$ with no theta term. 
Let $\mb{n}= (n^1, \dots,n^N)$ be the NLSM field. Again, the NLSM field is subject to the constraint 
$\mb{n}\cdot\mb{n}= n^a n_a = 1$. The Lagrangian in one spacetime dimension is
\beq
	\mathcal{L}= \frac{1}{2f}(\pd_t n^a)(\pd_t n_a)\ . \label{eq:ONlag}
\eeq
The canonical momentum conjugate to $n^a$ is 
$\pi_a= \frac{\pd \mathcal{L}}{\pd (\pd_t n^a)}= \frac{1}{f}(\pd_t n_a)$ and the Hamiltonian takes the form
\beq
	H= \frac{f}{2} \pi^a\pi_a\ . \label{eq:ON-ham}
\eeq 

In one spacetime dimension the analysis of constrained Hamiltonian systems from Sec.~\ref{sec:NLSM-quantization}
leads to the commutation relations 
\begin{subequations}
\label{eq:commutators}
\beqa
	\left[\hat{n}^a, \hat{n}^b\right] &=& 0  \\
	\left[\hat{n}^a, \hat{\pi}_b\right] &=& i\left({\delta^a}_b - \frac{\hat{n}^a \hat{n}_b}{\hat{r}^2}\right) \\
	\left[\hat{\pi}_a, \hat{\pi}_b\right] &=& \frac{i}{\hat{r}^2}(\hat{\pi}_a \hat{n}_b - \hat{\pi}_b \hat{n}_a)\ ,
\eeqa
\end{subequations}
where $\hat{r}^2 = \hat{n}^a\hat{n}_a$. The operator $\hat{r}^2$ commutes with all other operators by construction 
and so it can be set equal to one at this point, exactly as in Sec.~\ref{sec:NLSM-quantization}. However, we find it more convenient 
for the exposition in this appendix to leave this operator in place and only set it to one at the end of the analysis. 

The Schroedinger representation used in this paper for the NLSM commutation relations can be adapted to the case where the 
operator $\hat{r}^2$ is kept in the commutation relations. In this case the operator $\hat{n}^a$ again acts as multiplication by the 
coordinate $n^a$, but the momentum operator $\hat{\pi}_a$ now takes the form 
\beqa
	\hat{\pi}^a = -i \left({\delta_a}^b - \frac{n_a n^b}{r^2}\right)\frac{\pd}{\pd n^b}\ .
\eeqa
In this representation the quantity $\hat{\pi}^a\hat{\pi}_a$ appearing in the Hamiltonian takes the explicit form
\beq
	\hat{\pi}^a\hat{\pi}_a = -\left\{  \left(\delta^{ab} -  \frac{n^a n^b}{r^2}\right) \frac{\pd^2}{\pd n^a \pd n^b} - \frac{(N-1)}{r^2}n^a \frac{\pd}{\pd n^a} \right\}\ , \label{eq:momentum-squared}
\eeq
so that the Hamiltonian operator is
\beq
	\hat{H}= -\frac{f}{2}\left\{  \left(\delta^{ab} -  \frac{n^a n^b}{r^2}\right) \frac{\pd^2}{\pd n^a \pd n^b} - \frac{(N-1)}{r^2}n^a \frac{\pd}{\pd n^a} \right\}\ .
\eeq
To diagonalize the Hamiltonian we now prove the following statement.

\textit{Claim:} In the Schroedinger representation the squared sum of canonical momenta is related to $\Delta^{(N-1)}_{LB}$, 
the Laplace-Beltrami operator on the unit sphere $S^{N-1}$, by the equation
\beq
	\hat{\pi}^a\hat{\pi}_a= -\frac{1}{r^2}\Delta^{(N-1)}_{LB}\ .
\eeq

\textit{Proof:} Before imposing the NLSM constraint, the components $n^a$ of the NLSM field are coordinates on $\mathbb{R}^N$. 
The ordinary Laplacian on $\mathbb{R}^N$ is given in terms of the Laplace-Beltrami operator $\Delta^{(N-1)}_{LB}$ on $S^{N-1}$ 
by (see, e.g., Sec.~II.4 of Ref.~\cite{chavel1984eigenvalues})
\beq
	\frac{\pd^2}{\pd n^a \pd n_a} = \frac{\pd^2}{\pd r^2} + \frac{(N-1)}{r}\frac{\pd}{\pd r} + \frac{1}{r^2}\Delta^{(N-1)}_{LB}\ .
\eeq
Now using $\frac{\pd}{\pd r} = \frac{n^a}{r}\frac{\pd}{\pd n^a}$ and the relation
\beq
	\frac{\pd^2}{\pd r^2}= \frac{n^a n^b}{r^2}\frac{\pd^2}{\pd n^a \pd n^b}\ ,
\eeq
we find that
\beqa
	-\hat{\pi}^a\hat{\pi}_a &=& \left[\frac{\pd^2}{\pd n^a \pd n_a} - \left(\frac{\pd^2}{\pd r^2} + \frac{(N-1)}{r}\frac{\pd}{\pd r} \right)\right] \nnb \\
	&=& \frac{1}{r^2}\Delta^{(N-1)}_{LB}\ .
\eeqa
This completes the proof. $\blacksquare$

As we discussed earlier in this section, since the operator $\hat{r}^2$ commutes with all other operators, the NLSM constraint 
$\hat{r}^2=1$ can be enforced at any time in the correctly quantized theory. At this point we can then set $r^2=1$ to obtain the
final form for the Hamiltonian of the $O(N)$ NLSM,
\beq
	\hat{H}= -\frac{f}{2}\Delta^{(N-1)}_{LB}\ .
\eeq
We note here that in more careful approaches to the quantization of this model the Hamiltonian operator takes the form
$\hat{H}= \frac{f}{2}\left(-\Delta^{(N-1)}_{LB} + \overline{E}(N)\right)$, where $\overline{E}(N)$ is a constant
shift of the energy depending only on $N$ (although in the literature there is still some disagreement about the correct value of
$\overline{E}(N)$). In the straightforward approach used in this appendix this shift is not present. 
 
The spectrum of $\Delta^{(N-1)}_{LB}$, as well as its eigenfunctions and their multiplicities, can be found in standard
references on Riemannian geometry, for example Ref.~\cite{chavel1984eigenvalues}. Using these standard results 
we find that the eigenvalues of $\hat{H}$ are labeled by a positive integer $\ell$ and given explicitly by
\beq
	E_{\ell}= \frac{f}{2} \ell (\ell + N -2)\ ,\ \ell \in \mathbb{N}\ , \label{eq:ON-spectrum}
\eeq
in agreement with previous results on the spectrum of this model. 
The ground state of this theory has energy zero (or $\frac{f}{2}\overline{E}(N)$ for a non-zero shift in the energies), and the
difference between the energy of the ground state and first excited state is given by
\beq
	m \equiv E_1-E_0= \frac{f}{2} (N-1)\ .
\eeq
To get a complete understanding of the $O(N)$ NLSM in one spacetime dimension, 
we now give a full analysis of the symmetries of this system. 

\subsection{Symmetry analysis}

Consider the Lie algebra $so(N)$ of the Lie group $SO(N)$. In the fundamental (i.e., $N\times N$) representation, 
one possible basis of the Lie algebra consists of the anti-symmetric $N\times N$
matrices $\mb{E}^{ij}$ which contain a $1$ in the $(i,j)$ entry, a $-1$ in the $(j,i)$ entry, and zero in all other entries. 
Since $\mb{E}^{ij}=-\mb{E}^{ji}$, and since $i$ and $j$ must be distinct for this to make sense,  
we arrive at the correct number $N(N-1)/2$ of generators of $so(N)$. It is also convenient to define the matrices
$\mb{E}^{ii}$ for any $i$ to be equal to the matrix with all entries equal to zero.
The matrix elements of $\mb{E}^{ij}$ are
\beq
	{(\mb{E}^{ij})^a}_b= \delta^{ia}{\delta^j}_b- \delta^{ja}{\delta^i}_b\ .
\eeq
This definition also works when $i=j$ and yields the zero matrix in that case.

The Lagrangian of Eq.~\eqref{eq:ONlag} has an $SO(N)$ global symmetry which is reflected in the fact that it is invariant 
under the transformation $n^a \to {R^a}_b n^b$ for any $O(N)$ matrix $R$. The infinitesimal form of this transformation
is $n^a \to n^a + \delta n^a$ with $\delta n^a = \ep {(\mb{E}^{ij})^a}_b n^b$, for a small constant $\ep$. By 
making $\ep$ time-dependent we derive the conserved currents of this model,
\beq
	J_{ab}= \frac{1}{f}(\pd_t n_a n_b - \pd_t n_b n_a)\ .
\eeq
Since we are in $0+1$ dimensions the conserved charge operators are obtained from this current simply by replacing
$\pd_t n_a$ with $\hat{\pi}_a$, so the conserved charge operators are (note that we have chosen a particular operator
ordering here)
\beq
	\hat{Q}_{ab}= \hat{\pi}_a \hat{n}_b - \hat{\pi}_b \hat{n}_a\ .
\eeq
The commutator of two momenta from Eq.~\eqref{eq:commutators} can be rewritten 
(at this point we set $\hat{r}^2=1$ in the commutators) in terms of these charge operators as 
\beq
	[\hat{\pi}_a, \hat{\pi}_b] = i\hat{Q}_{ab}\ .
\eeq
In the Schrodinger representation (with $r^2=1$), the charges $\hat{Q}_{ab}$ take the simple form
\beq
	 \hat{Q}_{ab}= i (n_a \frac{\pd}{\pd n^b} - n_b \frac{\pd}{\pd n^a})\ .
\eeq
When acting on functions of $n^a$ these derivative operators obey the Lie algebra of $so(N)$,
\beq
	[ \hat{Q}_{ab}, \hat{Q}_{cd}]= \delta_{ac} \hat{Q}_{bd} - \delta_{ad} \hat{Q}_{bc} + \delta_{bd} \hat{Q}_{ac} - \delta_{bc} \hat{Q}_{ad}\ . \label{eq:soN-Lie-algebra}
\eeq 

We now show that the Hamiltonian Eq.~\eqref{eq:ON-ham} of this system is proportional to the quadratic Casimir of 
$so(N)$. It then follows that the problem of diagonalizing the Hamiltonian of the $O(N)$ NLSM reduces to a study of the
representation theory of $so(N)$, which is already well-known. The quadratic Casimir of $so(N)$ is given by the sum of the 
squares of all the generators $\hat{Q}_{ab}$. We know that half
of these are redundant since $\hat{Q}_{ab}= - \hat{Q}_{ba}$, but we can exploit this fact and the fact that 
$\hat{Q}_{aa}=0$ to write the Casimir as simply
\beq
	\hat{\mathcal{C}}= \frac{1}{2}\hat{Q}_{ab}\hat{Q}^{ab}\ ,
\eeq
where we have summed over all values of $a$ and $b$ with no restrictions. By explicit computation one can show that
$\hat{\mathcal{C}}= \hat{\pi}^a\hat{\pi}_a$ (when we set $r^2=1$ in Eq.~\eqref{eq:momentum-squared}), 
and so the Hamiltonian can be re-written as
\beq
	\hat{H}= \frac{f}{2}\hat{\mathcal{C}}\ .
\eeq
In this form one can clearly see the relationship between the Hamiltonian and the $SO(N)$ symmetry of this model.

\section{Regularization of the NLSM Hamiltonian}
\label{app:reg}

In Sec.~\ref{sec:wavefcn-gap} we studied the energy gap of the NLSM in the disordered ($f\to\infty$) limit and 
at $\theta=0$ using a lattice regularization. We briefly indicated there that some kind of regularization scheme was necessary to 
study the excited states of the NLSM, and then we immediately implemented the lattice regularization. In this appendix we explain in 
detail why it is necessary to regularize the NLSM Hamiltonian to study the excited states, and we also discuss an alternative 
regularization for the theory on flat space which does not use a lattice. We show that this alternative regularization gives results for 
the energy gap of the theory which are consistent with the result coming from the lattice regularization. Based on this evidence we 
expect that any sensible regularization scheme will give a result for the energy gap of the NLSM which agrees with our result 
computed using the lattice regularization. In this appendix we focus on the NLSM Hamiltonian at $\theta=0$. As we explained in 
Sec.~\ref{sec:wavefcn-gap}, the NLSM Hamiltonian at $\theta=2\pi k$, $k\in\mathbb{Z}$, is related to the NLSM Hamiltonian
at $\theta=0$ by a unitary transformation. Therefore any result on the spectrum of this theory at $\theta=0$ will also hold
for the theory at $\theta=2\pi k$ for integer $k$.

We start by explaining why the NLSM Hamiltonian must be regularized before excited states can be constructed. 
Recall that in the limit of large coupling $f$, the Hamiltonian for the $O(N)$ NLSM in $D$ spatial dimensions takes the form
\beq
	\hat{H}= \frac{f}{2}\int d^D \mb{x}\ \hat{\pi}_a(\mb{x})\hat{\pi}^a(\mb{x})\ ,
\eeq
where $\hat{\pi}_a(\mb{x})$ takes the form shown in Eq.~\eqref{eq:schro-rep} in the Schroedinger representation used in 
this paper. The ground state of this Hamiltonian has zero energy and is characterized by the property that it is annihilated by 
$\hat{\pi}^a(\mb{x})$ for each $a$. Therefore, to construct the ground state we only have to consider the action of a single 
operator $\hat{\pi}^a(\mb{x})$ on functionals of the NLSM field. On the other hand, to construct excited states we need
to act with the product $\hat{\pi}_a(\mb{x})\hat{\pi}^a(\mb{x})$. This operator is not well-defined in the NLSM field theory, as we now show. 


To see the problem with the operator $\hat{\pi}_a(\mb{x})\hat{\pi}^a(\mb{x})$, we look at the action of 
$\hat{\pi}_a(\mb{x})\hat{\pi}^a(\mb{y})$ on some functional $F$ of the NLSM field. A short calculation shows that
\begin{widetext}
\begin{align}
	\hat{\pi}_a(\mb{x})\hat{\pi}^a(\mb{y})F =  (N-1)\delta^{(D)}(\mb{x}-\mb{y}) n^a(\mb{x})\frac{\delta F}{\delta n^a(\mb{x})} + n^a(\mb{y})n^b(\mb{y})\frac{\delta^2 F}{\delta n^a(\mb{x})\delta n^b(\mb{y})} -n_a(\mb{x})n^b(\mb{x})n^a(\mb{y})n^c(\mb{y})\frac{\delta^2 F}{\delta n^b(\mb{x})\delta n^c(\mb{y})}\ .
\end{align}
\end{widetext}
We see that the operator $\hat{\pi}_a(\mb{x})\hat{\pi}^a(\mb{y})$ will diverge as $\mb{y}$ approaches $\mb{x}$ because of the
presence of the delta function in the first term of this expression. This contact divergence implies that the
product $\hat{\pi}_a(\mb{x})\hat{\pi}^a(\mb{x})$ of momentum operators at the same point $\mb{x}$ in space is ill-defined, and
this is the basic reason why some regularization scheme is needed to construct excited states in this field theory. 

Since the divergence in the operator $\hat{\pi}_a(\mb{x})\hat{\pi}^a(\mb{x})$ is due to the fact that both factors of 
$\hat{\pi}_a(\mb{x})$ are evaluated at the same point, i.e., the problem is associated with short distances, one way to 
regulate the operator is to discretize space by introducing a lattice. This is exactly the approach we took in 
Sec.~\ref{sec:wavefcn-gap}. However, other regularization schemes are also possible and should give expressions for the energy 
gap which agree with the answers obtained from the lattice regularization. To show this we now discuss one alternative regularization 
scheme, in which we still consider the theory on a continuous space, but we introduce some non-locality to regulate the product 
$\hat{\pi}_a(\mb{x})\hat{\pi}^a(\mb{x})$. In this regularization scheme we first rewrite the Hamiltonian as
\beq
	\hat{H}= \frac{f}{2}\int d^D \mb{x}d^D \mb{y}\ \hat{\pi}_a(\mb{x})\hat{\pi}^a(\mb{y})\delta^{(D)}(\mb{x}-\mb{y})\ ,
\eeq
and then replace the delta function $\delta^{(D)}(\mb{x}-\mb{y})$ with any known regularized expression for a delta function,
\beq
	\delta^{(D)}(\mb{x}-\mb{y}) \to \eta^{(D)}_{\ep}(\mb{x}-\mb{y})\ .
\eeq
Here $\eta^{(D)}_{\ep}(\mb{x}-\mb{y})$ is some function of $\mb{x}-\mb{y}$ which has the property that
\beq
	\lim_{\ep\to 0} \int d^D\mb{x}\ f(\mb{x}) \eta^{(D)}_{\ep}(\mb{x}-\mb{y})= f(\mb{y})\ ,
\eeq
for any test function $f(\mb{x})$. The parameter $\ep$ has units of length, and it is this small parameter which serves as a 
regulator for the theory in this regularization scheme. We consider a concrete example of a such a function 
$\eta^{(D)}_{\ep}(\mb{x}-\mb{y})$ later in this appendix. In terms of this function the regularized Hamiltonian takes the form
\beq
	\hat{H}(\ep)= \frac{f}{2}\int d^D \mb{x}d^D \mb{y}\ \hat{\pi}_a(\mb{x})\hat{\pi}^a(\mb{y})\eta^{(D)}_{\ep}(\mb{x}-\mb{y})\ .
\eeq
This regularization scheme clearly introduces some non-locality into the theory, since the regularized Hamiltonian 
(with non-zero $\ep$) contains terms which involve the fields $\hat{\pi}_a$ at two different points in space.

Within this alternative regularization scheme we can also compute the energy gap between the ground state of the system and
the first excited state. The vacuum state of the NLSM in the large $f$ limit (and at $\theta=0$) is the constant wave functional
$\Psi[\mb{n}]=1$, which transforms in the trivial representation of $SO(N)$. Experience with the lattice regularization of this
theory, and intuition about the role of the $SO(N)$ symmetry in this problem, suggests that the lowest energy states should 
transform in the vector representation of $SO(N)$. The simplest such states are given by functionals of the form
\beq
	F^a[\mb{n}]= \int d^D \mb{x}\ n^a(\mb{x})F(\mb{x})\ ,
\eeq
where $F(\mb{x})$ is some arbitrary function of space. For example, we can construct a localized excitation by choosing
$F(\mb{x})$ to be localized in some region of space. Applying the regularized Hamiltonian to this state gives
\beq
	\hat{H}(\ep)F^a[\mb{n}]= m(\ep) F^a[\mb{n}]\ ,
\eeq
where the energy $m(\ep)$ in this regularization is given by
\beq
	m(\ep)= \frac{f}{2} (N-1) \eta^{(D)}_{\ep}(0)\ .
\eeq
At this point it is instructive to make a particular choice of regularization of the delta function. We choose the Poisson kernel,
\beq
	\eta^{(D)}_{\ep}(\mb{x}-\mb{y}) = \prod_{j=1}^D \frac{1}{\pi}\frac{\ep}{\ep^2 + (x^j-y^j)^2}\ ,
\eeq
but other choices are also possible (e.g., a heat kernel, etc.). If we also set $N= D+2$, which is the case of interest in this
paper for constructing NLSMs with theta term in $D+1$ spacetime dimensions, then we find that the mass gap in this regularization is
\beq
	m(\ep)= \frac{f}{2} \frac{D+1}{(\pi \ep)^D}\ .
\eeq
This answer is clearly consistent with the expression Eq.~\eqref{eq:mass-gap-flat-space} obtained from the lattice regularization,
and the two expressions coincide if we choose the lattice spacing $a$ to be related to the parameter $\ep$ via $a= \pi\ep$.
Therefore we expect that any sensible regularization of the NLSM Hamiltonian will give results consistent with those that
we derived in Sec.~\ref{sec:wavefcn-gap} using the lattice regularization.

\section{Symplectic geometry approach to Hamiltonian formalism for field theories}
\label{app:symplectic}

In this appendix we review the symplectic geometry approach to the Hamiltonian dynamics of a classical field theory. 
This is essentially a ``functional" version of what one does in the symplectic geometry approach to a classical dynamical system
with finitely many degrees of freedom (a review of the latter for physicists is given in Ch.~11 of  Ref.~\cite{stonebook}).
We apply this formalism in Sec.~\ref{sec:NLSM-quantization-curved} to determine the correct form of the Poisson 
bracket in the theory of a free scalar field on a $D$-dimensional curved space $\mathcal{M}$ with some Riemannian
metric $G_{ij}$, and then we use this information to quantize the $O(D+2)$ NLSM with theta term on the curved space 
$\mathcal{M}$. One of the main advantages of the symplectic geometry approach is that it provides a formalism which one can rely 
on to understand the classical dynamics, and in particular the correct form of the Poisson bracket, in systems which cannot be
analyzed by conventional methods more familiar to physicists. The correct form of the Poisson brackets is essential for
quantization, and so this method is  useful for the proper quantization of an unfamiliar system.
This material is standard in the field theory literature. Therefore, in this section we simply give a summary of
this material in the infinite-dimensional setting in exact analogy to the development on a finite-dimensional phase space as found, for
example, in Ref.~\cite{stonebook}. In addition, we note here that a very similar infinite-dimensional symplectic geometry approach
is used in establishing the equivariant localization formulas for phase space path integrals in quantum mechanics (see, for example, 
Sec.~4.3 of Ref.~\cite{szabo2003equivariant}).

To start, consider a field theory on the $D$-dimensional space $\mathcal{M}$, and let 
$\{\Phi^a(\mb{x})\}_{\mb{x}\in \mathcal{M}}$ (for some range of the index $a$) denote the coordinates on the 
infinite-dimensional phase space for the system under consideration. In the example of a free scalar field $\phi(\mb{x})$
 we could choose $\Phi^1(\mb{x})= \phi(\mb{x})$, $\Phi^2(\mb{x})= \pi(\mb{x})$, where $\pi(\mb{x})$ is the momentum
conjugate to $\phi(\mb{x})$, but in general (as in the finite-dimensional case) it is not 
necessary to have a definite decomposition into ``coordinates" and ``momenta". 
In fact, in many cases it is \emph{impossible} to find
a definition of coordinates and momenta which is valid on the entire phase space. We use the notation 
$\vec{\Phi}(\mb{x})= (\Phi^1(\mb{x}),\Phi^2(\mb{x}),\dots)$ to denote the collection of all field variables 
$\Phi^a(\mb{x})$ at the single point $\mb{x}$. The functional derivatives $\frac{\delta}{\delta\Phi^a(\mb{x})}$ with 
respect to the phase space coordinates form a basis
of the tangent space at a point in this phase space. We also introduce the coordinate differentials $\delta \Phi^a(\mb{x})$,
which form a basis for the cotangent space at a point in phase space. We have the natural pairing 
between the basis elements of the tangent and cotangent spaces,
\beq
	\delta\Phi^a(\mb{x}) \left( \frac{\delta}{\delta\Phi^b(\mb{y})}\right)= {\delta^a}_b \delta^{(D)}(\mb{x}-\mb{y})\ .
\eeq
On phase space we also introduce an exterior derivative $\delta$ which acts on any functional $F$ of
the phase space coordinates as
\beq
	\delta F= \int d^D \mb{x}\ \frac{\delta F}{\delta \Phi^a(\mb{x})} \delta\Phi^a(\mb{x})\ .
\eeq
The wedge product of differentials $\delta\Phi^a(\mb{x})$ is defined in the usual way by
\beq
	\delta\Phi^a(\mb{x})\wedge \delta\Phi^b(\mb{y}) = \delta\Phi^a(\mb{x}) \otimes \delta\Phi^b(\mb{y}) - \delta\Phi^b(\mb{y}) \otimes \delta\Phi^a(\mb{x})
\eeq

A general $p$-form $\al$ on phase space has the form
\begin{widetext}
\beq
	\al = \frac{1}{p!}\int \left(\prod_{j=1}^p d^D \mb{x}_j \right) \al_{a_1\cdots a_p}\left[\vec{\Phi}(\mb{x}_1),\dots,\vec{\Phi}(\mb{x}_p);\mb{x}_1,\dots,\mb{x}_p\right]\ \delta\Phi^{a_1}(\mb{x}_1)\wedge\cdots\wedge\delta\Phi^{a_p}(\mb{x}_p)\ ,
\eeq
\end{widetext}
where $\al_{a_1\cdots a_p}\left[\vec{\Phi}(\mb{x}_1),\dots,\vec{\Phi}(\mb{x}_p);\mb{x}_1,\dots,\mb{x}_p\right]$ are the 
components of $\al$. The notation is meant to indicate that the components of $\al$ can depend on the fields $\Phi^a(\mb{x}_j)$
at the coordinates $\mb{x}_j$, and they can also depend explicitly on the coordinates $\mb{x}_j$. The action of the 
exterior derivative $\delta$ on $p$-forms is defined by the usual axioms: (i) $\delta^2 F= 0$ for any functional
$F$ on phase space, and (ii) $\delta(\al \wedge \beta)= \delta\al\wedge\beta +(-1)^p \al\wedge\delta\beta$ for any $p$-form $\al$ and 
any $q$-form $\beta$ (i.e., $\delta$ is an antiderivation). A general vector field $\un{V}$ on the phase space has the form
\beq
	\un{V}= \int d^D\mb{x}\ V^a\left[\vec{\Phi}(\mb{x});\mb{x}\right] \frac{\delta}{\delta\Phi^a(\mb{x})}\ ,
\eeq
where $V^a\left[\vec{\Phi}(\mb{x});\mb{x}\right]$ are the components of $\un{V}$.
The \emph{interior multiplication} of a form $\al$ by a vector $\un{V}$, denoted $i_{\un{V}} \al$, is given by
\begin{widetext}
\beq
	i_{\un{V}} \al= \frac{1}{(p-1)!}\int \left(\prod_{j=1}^p d^D \mb{x}_j \right) V^a\al_{a a_2\cdots a_p}\ \delta\Phi^{a_2}(\mb{x}_2)\wedge\cdots\wedge\delta\Phi^{a_p}(\mb{x}_p)\ ,
\eeq
\end{widetext}
where we suppressed the arguments of $\al_{{aa_2\cdots a_p}}$ and $V^a$ for brevity. 

After these preliminaries we are now ready to develop the canonical formalism on this infinite-dimensional phase space in exact
analogy to the development on a finite-dimensional phase space (see, for example, Ch.~11 of Ref.~\cite{stonebook}).
First, we introduce a symplectic form $\Omega$ on phase space, whose components are defined by
\begin{align}
	\Omega= \frac{1}{2}\int d^D\mb{x}_1 d^D\mb{x}_2\ \Omega_{ab}\left[\vec{\Phi}(\mb{x}),\vec{\Phi}(\mb{y});\mb{x},\mb{y}\right]\  \delta\Phi^a(\mb{x})\wedge\delta\Phi^b(\mb{y})\ .
\end{align}
As usual, we require that $\Omega$ is closed, $\delta\Omega=0$. Note also that in this infinite-dimensional case, the components
$\Omega_{ab}\left[\vec{\Phi}(\mb{x}),\vec{\Phi}(\mb{y});\mb{x},\mb{y}\right]$ of $\Omega$ only need to be antisymmetric 
under the \emph{simultaneous} exchange $a \leftrightarrow b$ and $\mb{x} \leftrightarrow \mb{y}$. Next, for 
any functional $F$ on phase space we define a vector field $\un{V}_F$ by the relation 
\beq
	\delta F= - i_{\un{V}_F}\Omega\ . \label{eq:vector-field}
\eeq
The reason for defining these vector fields in this way is that they allow for a coordinate-independent definition of the 
Poisson bracket of two functionals $F_1$ and $F_2$ on the phase space.
The Poisson bracket for $F_1$ and $F_2$ is given in terms of the corresponding vector fields 
$\un{V}_{F_1}$ and $\un{V}_{F_2}$ by\footnote{We use an opposite sign in this equation as compared with
Ref.~\cite{stonebook}.}
\beq
	\{F_1,F_2\}=  i_{\un{V}_{F_1}} i_{\un{V}_{F_2}}\Omega\ . \label{eq:PB}
\eeq
Finally, Hamilton's equations are equivalent to the single equation
\beq
	\delta H= - i_{\un{V}_H}\Omega\ , \label{eq:Hamilton-eqns}
\eeq
where $\un{V}_H$ is the vector field whose components are the time derivatives of phase space coordinates,
\beq
	\un{V}_H =\int d^D\mb{x}\ \dot{\Phi}^a(\mb{x}) \frac{\delta}{\delta\Phi^a(\mb{x})}\ ,
\eeq
and where the dot represents a time derivative, $\dot{(\ )} := \frac{d}{dt}(\ )$.

\section{Intrinsic construction of the Wess-Zumino term for $D=1$}
\label{app:WZ-intrinsic}

In this appendix we explain in detail a simple example of the intrinsic construction of the Wess-Zumino (WZ) term which
we mentioned in Sec.~\ref{sec:NLSM-quantization-curved}. 
This intrinsic construction is the method of choice for constructing the WZ term when other standard
constructions fail, for example in the case where the spatial manifold $\mathcal{M}$ does not admit a spin structure and 
also cannot be realized as the boundary of any higher-dimensional manifold $\mathcal{B}$. The details of the intrinsic construction 
are, however, more complicated than the more standard constructions. For this reason we only present the simplest example of
the construction, which is the case where $\mathcal{M}$ has dimension $D=1$, but this intrinsic construction is available in all dimensions. 
The ideas behind this construction date back to work of Wu and Yang in Ref.~\cite{wu1976dirac} and were formalized
by Alvarez in Ref.~\cite{alvarez1985topological} using the language of \v{C}ech cohomology. 
The basic idea is to write the WZ term as a sum of integrals over the space $\mathcal{M}$ of forms which are defined only locally in 
certain coordinate patches on the target manifold $\mathcal{T}$ of the NLSM. This sum of integrals over $\mathcal{M}$ is then 
supplemented with additional terms which account for the transition functions which are needed to go between coordinate patches on 
the target manifold. We also mention here that the methods of Ref.~\cite{alvarez1985topological} were used in
Ref.~\cite{polychronakos1987topological} to give an intrinsic definition of the Abelian Chern-Simons term on a 
three-dimensional manifold. 

We now present the intrinsic construction of the WZ term for the case where the spatial manifold $\mathcal{M}$ has dimension
$D=1$. Our discussion in this appendix applies only to the specific case where the target manifold of the NLSM is the
sphere $S^2$. Other two-dimensional target spaces may require more coordinate patches to be covered properly. 
As we discussed in the main sections of this paper, we assume that $\mathcal{M}$ is closed, oriented, and connected. 
For the case $D=1$ this implies that $\mathcal{M}$ is diffeomorphic to a circle. We take $x^1 \in [a,b)$ to be the coordinate
on this circle, and the NLSM field is taken to obey periodic boundary conditions $\mb{n}(a)=\mb{n}(b)$. For $D=1$ we have
an $O(3)$ NLSM, and for the purpose of constructing the WZ term it will be convenient to parametrize the field variables $n^a$, 
$a=1,2,3$, using spherical coordinates $\Phi$ and $\Theta$ as
\bseq
\beqa
	n^1 &=& \cos(\Phi)\sin(\Theta) \\
	n^2 &=& \sin(\Phi)\cos(\Theta) \\
	n^3 &=& \cos(\Theta)\ .
\eeqa
\eseq
In these coordinates the volume form on $S^2$ (the target manifold of the NLSM) takes the form 
\beq
	\omega_2= \sin(\Theta)d\Theta \wedge d\Phi\ .
\eeq
The sphere $S^2$ can be covered by two coordinate patches $U_N$ and $U_S$, defined in spherical coordinates as 
$U_N= \{(\Phi,\Theta) |\ \Phi\in[0,2\pi), \Theta\in[0,\pi-\Theta_0)\}$ and 
$U_S= \{(\Phi,\Theta) |\ \Phi\in[0,2\pi), \Theta\in(\Theta_0,\pi]\}$, for some fixed (perhaps small) angle $\Theta_0$. The patch
$U_N$ contains the north pole but not the south pole, and the patch $U_S$ contains the south pole but not the north pole.
On each patch the volume form can be expressed as a total derivative $\omega_2 = d\vth_N$ or $\omega_2= d\vth_S$, with
\bseq
\label{eq:symplectic-potentials-sphere}
\beqa
	\vth_N &=& (1-\cos(\Theta))d\Phi \\
	\vth_S &=& -(1+\cos(\Theta))d\Phi \ .
\eeqa
\eseq
On the intersection $U_S \cap U_N$ of the two coordinate patches we have $\vth_S - \vth_N= d\psi_{SN}$, where (up to an 
arbitrary constant)
\beq
	\psi_{SN} = -2\Phi\ . \label{eq:transition-function}
\eeq

At this point we recall that the ordinary construction of the WZ term for the NLSM on $\mathcal{M}$ uses an 
extended manifold $\mathcal{B}$ with $\pd\mathcal{B}=\mathcal{M},$ and an extension $\tilde{\mb{n}}$ of the NLSM field into 
$\mathcal{B}$ such that $\tilde{\mb{n}}|_{\pd\mathcal{B}}=\mb{n}$. In this case the standard construction for the WZ term is
\beq
	S_{WZ}[\mb{n}]= \frac{2\pi k}{\A_2}\int_{\mathcal{B}}\tilde{\mb{n}}^*\omega_2\ .
\eeq
Using the spherical coordinates $\Theta$ and $\Phi$ we find that the variation of the WZ term obtained from this standard 
construction is
\beq
	\delta S_{WZ}[\mb{n}]= \frac{2\pi k}{\A_2} \int_a^b dx^1\ \sin(\Theta) \left(\delta\Theta \pd_{1}\Phi - \delta\Phi \pd_{1}\Theta\right)\ , \label{eq:WZ-variation-spherical-coords}
\eeq
where $\pd_1 \equiv \frac{\pd}{\pd x^1}$.
We now present the intrinsic construction of the WZ term, which yields an expression for $S_{WZ}[\mb{n}]$ involving only 
integrations over the physical space $\mathcal{M}$, and gives the same formula for the variation with respect to the NLSM field.

The intrinsic construction takes as its starting point a geometric interpretation of the WZ term. The NLSM field $\mb{n}$ is a map from 
$\mathcal{M}$ to $S^2$, and the image of $\mathcal{M}$ under this map is a closed curve $\ell$ on $S^2$. The curve $\ell$ inherits
an orientation from the orientation of $\mathcal{M}$. On $S^2$ there exist 
regions $\mathcal{C}$ and $\mathcal{C}' = S^2\backslash \mathcal{C}$ such that $\pd \mathcal{C} = \ell$ and 
$\pd\mathcal{C}' = \overline{\ell}$, where $\overline{\ell}$
is the curve $\ell$ with the opposite orientation. Using this information we define the WZ term for the field configuration 
$\mb{n}$ using the \emph{signed} area of the regions $\mathcal{C}$ or $\mathcal{C}'$ as
\beq
	S_{WZ}[\mb{n}] = \frac{2\pi k}{\A_2}\text{Area}[\mathcal{C}]\ , \label{eq:WZ-term-as-area}
\eeq
or
\beq
	S'_{WZ}[\mb{n}] = -\frac{2\pi k}{\A_2}\text{Area}[\mathcal{C}']\ .
\eeq
The minus sign in the second equation is there to keep track of the fact that the boundary of $\mathcal{C}'$ is
$\overline{\ell}$, which has the opposite orientation of $\ell=\pd\mathcal{C}$. The reason for defining the WZ term in this way
is that with this definition we have
\beqa
	e^{iS_{WZ}[\mb{n}]} &=& e^{i\left(S_{WZ}[\mb{n}]-S'_{WZ}[\mb{n}]\right)}e^{iS'_{WZ}[\mb{n}]}  \nnb \\
	&=& e^{i2\pi k}e^{iS'_{WZ}[\mb{n}]}\ .
\eeqa
Therefore, we see that the exponential of the WZ term will be independent of the choice of $S_{WZ}[\mb{n}]$ or 
$S'_{WZ}[\mb{n}]$ as long as the level $k$ of the WZ term is quantized, $k\in\mathbb{Z}$, which is the usual result. In what follows 
we work with the first formula Eq.~\eqref{eq:WZ-term-as-area}. 
 
The formula Eq.~\eqref{eq:WZ-term-as-area} instructs us to integrate the volume form 
$\omega_2$ over the region $\mathcal{C} \subset S^2$. In the case where the curve $\ell$ is contained only in the coordinate 
patch $U_S$ on $S^2$, we have $\omega_2=d\vth_S$ and so the WZ term takes the simple form
\beqa
	S_{WZ}[\mb{n}] &=& \frac{2\pi k}{\A_2}\int_{\ell}\vth_s \nnb \\
	&=& \frac{2\pi k}{\A_2} \int_{\mathcal{M}} \mb{n}^*\vth_S \nnb \\
	&=& -\frac{2\pi k}{\A_2}\int_a^b dx^1 (1+\cos(\Theta))\pd_{1}\Phi\ ,
\eeqa
where we used the expression for $\vth_S$ from Eqs.~\eqref{eq:symplectic-potentials-sphere}. 
If instead the curve $\ell$ is contained only in $U_N$, then we have a similar 
expression for $S_{WZ}[\mb{n}]$ with $\vth_S$ replaced with $\vth_N$. Finally, there is the possibility that the curve $\ell$ crosses 
through both coordinate patches on $S^2$. In this case the expression for $S_{WZ}[\mb{n}]$ obtained from the
formula Eq.~\eqref{eq:WZ-term-as-area} is more complicated.

\begin{figure}[t]
  \centering
    \includegraphics[width= .5\textwidth]{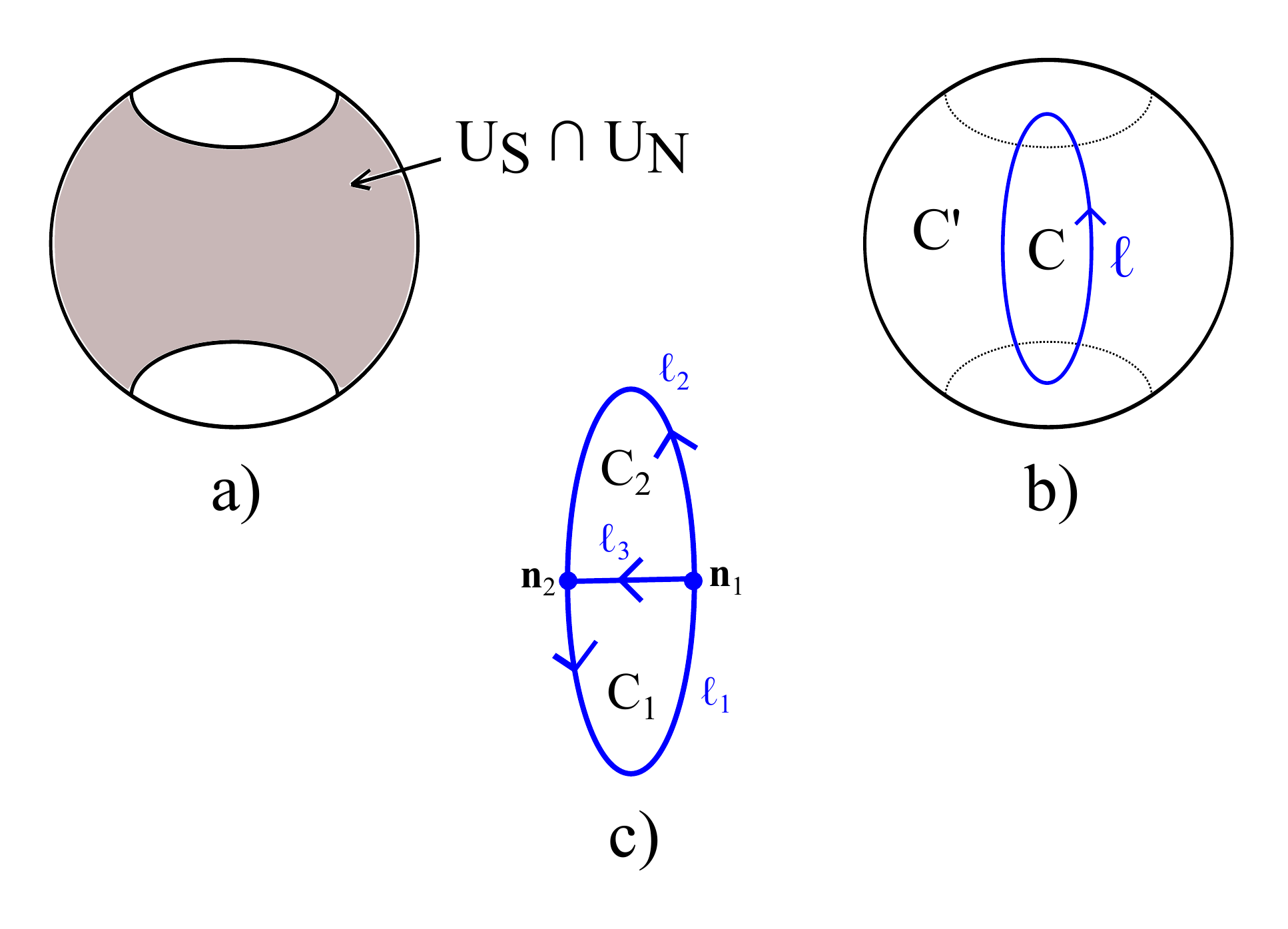} 
\vskip 10pt
 \caption{a) The shaded region shows the intersection $U_S \cap U_N$ of the two coordinate patches needed to cover
the entire sphere $S^2$. b) The curve $\ell$ and the regions $\mathcal{C}$ and $\mathcal{C}'$ whose union is the entire sphere 
$S^2$. The curve $\ell$ does not lie in a single coordinate patch $U_S$ or $U_N$, as can be seen from the dotted lines indicating the
boundary of the intersection $U_S\cap U_N$. c) A partition of the region $\mathcal{C}$ into two parts $\mathcal{C}_1$ and 
$\mathcal{C}_2$ using an additional curve $\ell_3$ which starts at the point $\mb{n}_1$ and ends at the point $\mb{n}_2$. 
The part $\mathcal{C}_1$ lies entirely in $U_S$, while the part $\mathcal{C}_2$ lies entirely in $U_N$.}
\label{fig:WZ-setup}
\end{figure}

To construct the WZ term in the case where the curve $\ell$ passes through both coordinate patches on $S^2$, we do the
following. First, we pick two points $s_1$ and $s_2$ on the interval $[a,b)$ such that $\mb{n}_{1}\equiv \mb{n}(s_1)$ and
$\mb{n}_2\equiv \mb{n}(s_2)$ lie on opposite sides of the curve $\ell$ in the region $U_S \cap U_N$. Next, we divide the curve 
$\ell$ into two pieces $\ell_1$ and $\ell_2$ such that $\ell=\ell_1+\ell_2$, where the sum is the composition of oriented 1-chains. 
Finally, we add a third curve $\ell_3$ which connects the points $\mb{n}_1$ and $\mb{n}_2$ by cutting through the region 
$U_S \cap U_N$, and we choose the orientation of this curve such that it is directed towards $\mb{n}_2$. The curve $\ell_3$ 
also divides the region $\mathcal{C}$ into two portions $\mathcal{C}_1$ and $\mathcal{C}_2$. We choose the points $s_1$ and $s_2$,
and also the curve $\ell_3$, so that $\mathcal{C}_1$ lies entirely in $U_S$ and $\mathcal{C}_2$ lies entirely in $U_N$.
This situation is illustrated in Fig.~\ref{fig:WZ-setup}. In this case we can compute $\text{Area}[\mathcal{C}]$ as
\begin{align}
	\text{Area}[\mathcal{C}] &= \text{Area}[\mathcal{C}_1]+\text{Area}[\mathcal{C}_2] \nnb \\
	&= \int_{\ell_1+\ell_3}\vth_S + \int_{\ell_2-\ell_3}\vth_N \nnb \\
	&= \int_{\ell_1}\vth_S + \int_{\ell_2}\vth_N + \int_{\ell_3}(\vth_S-\vth_N) \nnb \\
	&= \int_{\ell_1}\vth_S + \int_{\ell_2}\vth_N + \psi_{SN}(\mb{n}_2) - \psi_{SN}(\mb{n}_1)\ ,  
\end{align}
where in the last step we used the equation $\vth_S -\vth_N = d\psi_{SN}$ on $U_S\cap U_N$. The integrals in this expression can
be pulled back to $\mathcal{M}$ to give a final expression for the WZ term in the form
\begin{align}
	S_{WZ}[\mb{n}] = \frac{2\pi k}{\A_2}\Bigg(&\int_{\mb{n}^{-1}(\ell_1)}\mb{n}^*\vth_S + \int_{\mb{n}^{-1}(\ell_2)}\mb{n}^*\vth_N \nnb \\ 
&+ \psi_{SN}(\mb{n}(s_2)) - \psi_{SN}(\mb{n}(s_1))   \Bigg)\ , \label{eq:WZ-final-intrinsic}
\end{align}
where $\mb{n}^{-1}(\ell_1)$ denotes the inverse image of the curve $\ell_1$ under the map $\mb{n}:\mathcal{M}\to S^2$, and likewise
for $\mb{n}^{-1}(\ell_2)$.

This form for the WZ term has the advantage that it only involves integrals over the physical space $\mathcal{M}$, and does
not require an extended space $\mathcal{B}$ or an extension $\tilde{\mb{n}}$ of the field configuration (it also does not
require a spin structure on $\mathcal{M}$). Therefore we 
refer to this construction of the WZ term as an \emph{intrinsic} construction. In addition, 
a short calculation shows that upon varying this expression with respect to the NLSM field, and using the explicit expression for
$\psi_{SN}$ from Eq.~\eqref{eq:transition-function}, the contributions from the points $s_1$ and $s_2$ cancel out. The end result 
for the variation of this form of the WZ term then turns out to be identical to Eq.~\eqref{eq:WZ-variation-spherical-coords}. 
Therefore we have succeeded in our original goal, which was to provide a construction of the WZ term for the NLSM theory on the 
manifold $\mathcal{M}$ which does not require a higher-dimensional manifold $\mathcal{B}$ with $\pd \mathcal{B}=\mathcal{M}$.
Also, we note here that adding an arbitrary constant to the transition function $\psi_{SN}$, which still gives a solution to the
equation $\vth_S -\vth_N =d\psi_{SN}$, will not change the final expression Eq.~\eqref{eq:WZ-final-intrinsic} for the 
WZ term since $\psi_{SN}$ appears in the WZ term in the combination 
$\psi_{SN}(\mb{n}(s_2)) - \psi_{SN}(\mb{n}(s_1))$.

\begin{figure}[t]
  \centering
    \includegraphics[width= .3\textwidth]{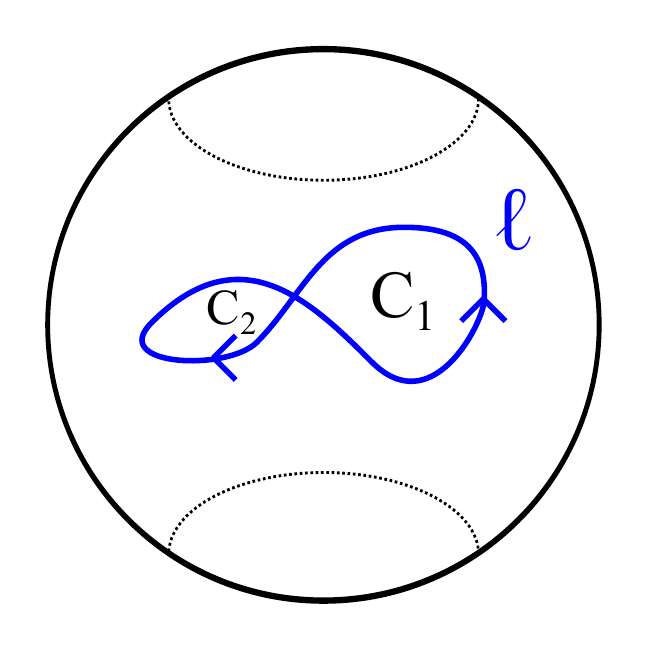} 
\vskip 10pt
 \caption{An example of a curve $\ell$ with a self-intersection 
arising from a map $\mb{n}: \mathcal{M} \to S^2$ which is not injective. In this
case the WZ term should be computed using the signed area $\text{Area}[\mathcal{C}_1] - \text{Area}[\mathcal{C}_2]$ of the
two regions $\mathcal{C}_1$ and $\mathcal{C}_2$ enclosed by the curve $\ell$, as shown in 
Eq.~\eqref{eq:WZ-term-self-intersection}. This is because the curve $\ell$ goes counterclockwise around the region 
$\mathcal{C}_1$ but clockwise around the region $\mathcal{C}_2$.}
\label{fig:self-intersection}
\end{figure}

Finally, we must discuss what one must do in the case that the curve $\ell$ on $S^2$ has self-intersections. The map 
$\mb{n}: \mathcal{M} \to S^2$ is not required to be an \emph{embedding} of $\mathcal{M}$ (i.e., $\mb{n}$ is not required
to be injective), so in general the curve $\ell$ can have self-intersections. In this case we again define the WZ term using the signed 
area of the regions enclosed by $\ell$. For example, for the situation shown in Fig.~\ref{fig:self-intersection} we define
\beq
	S_{WZ}[\mb{n}]= \frac{2\pi k}{\A_2}\left( \text{Area}[\mathcal{C}_1] - \text{Area}[\mathcal{C}_2]  \right)\ . 
\label{eq:WZ-term-self-intersection}
\eeq
The reason for defining the WZ term in this way is that we ultimately want $S_{WZ}[\mb{n}]$ to reduce to a line integral
of $\vth_S$ or $\vth_N$ along the curve $\ell$, modulo the addition of suitable constant terms in the intersection 
$U_S\cap U_N$ as discussed above. Since in the example in Fig.~\ref{fig:self-intersection} 
the curve $\ell$ wraps around the regions $\mathcal{C}_1$ and $\mathcal{C}_2$ in opposite directions, we attach opposite signs to 
the areas of these two regions in the definition of the WZ term so that $S_{WZ}[\mb{n}]$ reduces to a line integral of $\vth_S$ or 
$\vth_N$ along $\ell$.

\end{document}